\documentclass[submission,copyright,creativecommons]{eptcs}
\providecommand{\event}{SYNT 2014} 

\usepackage{xcolor,pgf,tikz,pgflibraryarrows,pgffor,pgflibrarysnakes}
\usetikzlibrary{automata,shapes,snakes,backgrounds,decorations.pathreplacing}
\usepackage{bbold}
\usepackage{lscape}
\usepackage{amssymb,amsmath,amsthm}
\usepackage{amssymb,stmaryrd}
\usepackage{algorithm}
\usepackage[noend]{algorithmic}
\usepackage{comment}
\usepackage{times}
\usepackage{url}
\usepackage{wrapfig}
\usepackage{placeins}
\usepackage{enumerate}
\usepackage{algorithm}
\usepackage{algorithmic}
\usepackage{multirow}
\usepackage{rotating}
\usepackage{wrapfig}

\algsetup{linenosize=\scriptsize}
\DeclareSymbolFont{cmmathcal}{OMS}{cmsy}{m}{n}
\DeclareSymbolFontAlphabet{\mathcal}{cmmathcal}

\usepackage{enumitem}
\setlist[1]{itemsep=-1pt}

\hypersetup{                    
    colorlinks=true,                
    breaklinks=true,                
    urlcolor= black,                 
    linkcolor= black,                
    citecolor= black                
    }

\input{env}

\title{
Symblicit algorithms for optimal strategy synthesis in monotonic Markov decision processes\thanks{This work has been partly supported by ERC Starting Grant (279499: inVEST), ARC project (number AUWB-2010-10/15-UMONS-3) and European project Cassting (FP7-ICT-601148). An extended version of this paper is available at~\cite{2014arXiv1402.1076B}.}}
\author{Aaron Bohy \quad\qquad V\'eronique Bruy\`ere
\institute{Universit\'e de Mons\\ Belgium}
\and
Jean-Fran\c{c}ois Raskin
\institute{Universit\'e Libre de Bruxelles\\ Belgium}}

\def\titlerunning{Symblicit algorithms for mean-payoff and shortest path in monotonic MDPs}
\def\authorrunning{A. Bohy, V. Bruy\`ere \& J.-F. Raskin}

\newcommand{\acaciaplus}{\ensuremath{\sf{Acacia+}}}
\newcommand{\prism}{\ensuremath{\sf{PRISM}}}
\newcommand{\modest}{\ensuremath{\sf{MODEST}}}
\newcommand{\mrmc}{\ensuremath{\sf{MRMC}}}
\newcommand{\quasy}{\ensuremath{\sf{QUASY}}}

\newcommand{\nat}{{\mathbb{N}}}
\definecolor{light-gray}{gray}{0.9}
\definecolor{light-gray2}{gray}{0.96}

\newcommand{\defeq}{\stackrel{\sf{def}}{=}}

\newcommand{\CTL}{\ensuremath{\sf{CTL}} }
\newcommand{\PCTL}{\ensuremath{\sf{PCTL}} }
\newcommand{\LTL}{\ensuremath{\sf{LTL}} }
\newcommand{\LTLMP}{\textnormal{\ensuremath{\sf{LTL_\textsf{MP}}}} }

\newcommand{\MP}{\mathsf{MP}}
\newcommand{\TS}{\mathsf{TS}}
\newcommand{\mpval}{\mathsf{Val}}

\newcommand{\probI}{\pi_I}

\newcommand{\R}{\mathbb{R}}

\newcommand{\Z}{\mathbb{Z}}

\newcommand{\distrset}{\mathcal{D}}

\newcommand{\domain}{\mathsf{Dom}}
\newcommand{\ActionsO}{\Sigma}
\newcommand{\ActionsI}{T}
\newcommand{\actionO}{\sigma}
\newcommand{\actionI}{\tau}
\newcommand{\edges}{\textnormal{\textbf{E}}}
\newcommand{\distr}{\textnormal{\textbf{D}}}
\newcommand{\probmat}{\textnormal{\textbf{P}}}

\newcommand{\id}{\textnormal{\textbf{I}}}
\newcommand{\totfunc}{\mathcal{F}_\textnormal{tot}}
\newcommand{\enabledactions}{\ActionsO_s}
\newcommand{\enabledactionsprime}{\ActionsO_{s'}}

\newcommand{\enabledstates}{S_\actionO}
\newcommand{\reward}{\textnormal{\textbf{C}}}
\newcommand{\ETP}{\mathbb{E}^{\textnormal{TS}_G}}
\newcommand{\EMP}{\mathbb{E}^\textnormal{MP}}
\newcommand{\Edot}{\mathbb{E}^{~\cdot}}

\newcommand{\symbMDP}{\mathcal{M}}
\newcommand{\symbReward}{\mathcal{C}}
\newcommand{\symbMC}{\symbMDP_{\lambda_n}}
\newcommand{\symbRewardMC}{\symbReward_{\lambda_n}}
\newcommand{\symbQuotient}{\symbMC'}
\newcommand{\symbRewardQuotient}{\symbRewardMC'}

\newcommand{\symbValue}{\mathcal{X}}
\newcommand{\symbGoal}{\mathcal{G}}
\newcommand{\explQuotient}{M_{\lambda_n}'}
\newcommand{\explRewardQuotient}{\reward_{\lambda_n}'}
\newcommand{\equivlump}{\sim_L}

\newcommand{\diff}{\backslash}
\newcommand{\pseudclos}{\updownarrow\!\!}
\newcommand{\antclos}{\downarrow\!\!}
\newcommand{\antunion}{\dot\cup}
\newcommand{\antinter}{\dot\cap}
\newcommand{\PA}{PA}

\newcommand{\monMDP}{M_\preceq}
\newcommand{\monMC}{M_{\preceq,\lambda}}

\newcommand{\PreMC}{\Pre_\lambda}
\newcommand{\Pre}{\textnormal{\textsf{Pre}}}

\newcommand{\val}{block}

\newcommand{\equivdistr}{\sim_{\distr,\lambda}}
\newcommand{\partdistr}{S_{\equivdistr}}
\newcommand{\equivreward}{\sim_{\reward,\lambda}}
\newcommand{\partreward}{S_{\equivreward}}

\newcommand{\equivsigmarew}{\sim_{\reward,\actionO}}
\newcommand{\equivsigmaprob}{\sim_{\probmat,\actionO}}

\newcommand{\equivstrat}{\sim_\lambda}

\newcommand{\equivstratp}{\sim_{\lambda'}}
\newcommand{\partstrat}{S_{\equivstrat}}

\newcommand{\partlump}{S_{\equivlump}}

\newcommand{\equiva}{\sim_{l_\actionO}}
\newcommand{\parta}{(S_{\actionO})_{\equiva}}
\newcommand{\equivc}{\sim_{\probmat,\actionO,C}}
\newcommand{\partc}{S_{\equivc}}

\newcommand{\equivalump}{\sim_{l_\actionO \wedge L}}
\newcommand{\partalump}{(S_\actionO)_{\equivalump}}
\newcommand{\partprom}{(S_\actionO)_{\equivalump}^<}

\newcommand{\laB}{l_\actionO(B)}
\newcommand{\las}{l_\actionO(s)}
\newcommand{\lasp}{l_\actionO(s')}
\newcommand{\lac}{l_\actionO(C)}
\newcommand{\lad}{l_\actionO(D)}

\newcommand{\memout}{\!\!{\bf>\!4000\!}}

\begin{document}
  \maketitle	

\begin{abstract} 
When treating Markov decision processes (MDPs) with large state spaces, using explicit representations quickly becomes unfeasible. Lately, Wimmer et al.\ have proposed a so-called symblicit algorithm for the synthesis of optimal strategies in MDPs, in the quantitative setting of expected mean-payoff. This algorithm, based on the strategy iteration algorithm of Howard and Veinott, efficiently combines symbolic and explicit data structures, and uses binary decision diagrams as symbolic representation. The aim of this paper is to show that the new data structure of pseudo-antichains (an extension of antichains) provides another interesting alternative, especially for the class of monotonic MDPs. We design efficient pseudo-antichain based symblicit algorithms (with open source implementations) for two quantitative settings: the expected mean-payoff and the stochastic shortest path. For two practical applications coming from automated planning and \LTL synthesis, we report promising experimental results w.r.t. both the run time and the memory consumption.
%
\end{abstract}

\section{Introduction}

Markov decision processes~\cite{putermanMDP,DBLP:books/daglib/0020348} (MDPs) are rich models that exhibit both nondeterministic choices and stochastic transitions. Model-checking and synthesis algorithms for MDPs exist for logical properties expressible in the logic \PCTL\cite{DBLP:journals/fac/HanssonJ94}, a stochastic extension of \CTL\cite{DBLP:conf/lop/ClarkeE81}, and are implemented in tools like \prism~\cite{KNP11}, \modest~\cite{DBLP:conf/fdl/Hartmanns12}, \mrmc~\cite{DBLP:journals/pe/KatoenZHHJ11}\dots There also exist algorithms for {\em quantitative properties} such as the long-run average reward (mean-payoff) or the stochastic shortest path, that have been implemented in tools like \quasy~\cite{DBLP:conf/tacas/ChatterjeeHJS11} and \prism~\cite{DBLP:conf/vmcai/EssenJ12}.


There are two main families of algorithms for MDPs. First, {\em value iteration} algorithms assign values to states of the MDPs and refine locally those values by successive approximations. If a fixpoint is reached, the value at a state $s$ represents a probability or an expectation that can be achieved by an optimal strategy that resolves the choices present in the MDP starting from $s$. This value can be, for example, the maximal probability to reach a set of goal states.  Second, {\em strategy iteration} algorithms start from an arbitrary strategy and iteratively improve the current strategy by local changes up to the convergence to an optimal strategy. Both methods have their advantages and disadvantages. Value iteration algorithms usually lead to easy and efficient implementations, but in general the fixpoint is not guaranteed to be reached in a finite number of iterations, and so only approximations are computed. On the other hand, strategy iteration algorithms have better theoretical properties as convergence towards an optimal strategy in a finite number of steps is usually ensured, but they often require to solve systems of linear equations, and so they are more difficult to implement efficiently.


When considering large MDPs obtained from high level descriptions or as the product of several components, explicit methods often exhaust available memory and are thus impractical. This is the manifestation of the well-known {\em state explosion problem}. In non-probabilistic systems, symbolic data structures like binary decision diagrams (BDDs) have been investigated~\cite{DBLP:journals/iandc/BurchCMDH92} to mitigate this phenomenon. For probabilistic systems, multi-terminal BDDs  (MTBDDs) are useful but they are usually limited to systems with around $10^{10}$ or $10^{11}$ states only~\cite{prismwebsite}. Also, as mentioned above, some algorithms for MDPs rely on solving linear systems, and there is no easy use of BDD like structures for implementing such algorithms.


Recently, Wimmer et al.~\cite{DBLP:conf/qest/WimmerBBHCHDT10} have proposed a method that {\em mixes} symbolic and explicit representations to efficiently implement the Howard and Veinott strategy iteration algorithm~\cite{howard1960dynamic,veinott1966finding} to synthesize optimal strategies for mean-payoff objectives in MDPs. Their solution is as follows. First, the MDP is represented and handled symbolically using MTBDDs. Second, a strategy is fixed symbolically and the MDP is transformed into a Markov chain (MC). To analyze this MC, a linear system needs to be constructed from its state space. As this state space is potentially huge, the MC is first reduced by {\em lumping}~\cite{kemeny1960jl,buchholz1994exact} (bisimulation reduction), and then a (hopefully) compact linear system can be constructed and solved. Solutions to this linear system allow to show that the current strategy is optimal, or to obtain sufficient information to improve it. A new iteration is then started. The main difference between this method and the other methods proposed in the literature is its {\em hybrid nature}: it is symbolic for handling the MDP and for computing the lumping, and it is explicit for the analysis of the reduced MC. This is why the authors of~\cite{DBLP:conf/qest/WimmerBBHCHDT10} have coined their approach {\em symblicit}.

\vspace{-0.2cm}
\paragraph{Contributions.} In this paper, we build on the symblicit approach described above. Our contributions are threefold. 
First, we show that the symblicit approach and strategy iteration can also be efficiently applied to the {\em stochastic shortest path} problem. We start from an algorithm proposed by Bertsekas and Tsitsiklis~\cite{bertsekas1996neuro} with a preliminary step of de Alfaro~\cite{DBLP:conf/concur/Alfaro99}, and we show how to cast it in the symblicit approach.
Second, we show that alternative data structures can be more efficient than BDDs or MTBDDs for implementing a symblicit approach, both for mean-payoff and stochastic shortest path objectives. In particular, we consider a natural class of MDPs with {\em monotonic properties} on which our alternative data structure is more efficient. For such MDPs, as for subset constructions in automata theory~\cite{DBLP:conf/cav/WulfDHR06,DBLP:conf/tacas/DoyenR07}, antichain based data structures usually behave better than BDDs. The application of antichains to monotonic MDPs requires nontrivial extensions: for instance, to handle the lumping step, we need to generalize existing antichain based data structures in order to be closed under negation. To this end, we introduce a new data structure called {\em pseudo-antichain}.
Third, we have implemented our algorithms and we show that they are more efficient than existing solutions on natural examples of monotonic MDPs. We show that monotonic MDPs naturally arise in probabilistic planning~\cite{blum2000probabilistic} and when optimizing controllers synthesized from \LTL specifications with mean-payoff objectives~\cite{DBLP:conf/tacas/BohyBFR13}.

\vspace{-0.2cm}
\paragraph{Structure of the paper.} In Section~\ref{sec:prelim}, we recall useful definitions and we introduce the notion of monotonic MDP. In Section~\ref{sec:si}, we recall strategy iteration algorithms for mean-payoff and shortest path objectives, and we present the symblicit version of those algorithms. We introduce the notion of pseudo-antichains in Section~\ref{sec:pa}, and we describe our pseudo-antichain based symblicit algorithms in Section~\ref{sec:paalgo}. In Section~\ref{sec:experiments}, we propose two applications of the symblicit algorithms and give experimental results. Finally in Section~\ref{sec:conclusion}, we summarize our results.

\section{Background and studied problems}\label{sec:prelim}

In this section, we recall useful definitions and we introduce the notion of monotonic Markov decision process. We also state the problems that we study. 

\paragraph{Stochastic models.}  
We denote by $\domain(f)$ the domain of definition of a function~$f$, and by $\totfunc(A, B)$ the set of total functions from $A$ to $B$. A \textit{probability distribution} over a finite set $A$ is a total function $\pi : A \rightarrow [0,1]$ such that $\sum_{a \in A} \pi(a) = 1$. 
We denote by $\distrset(A)$ the set of probability distributions over $A$. 

A \textit{discrete-time Markov chain (MC)} is a tuple $(S, \probmat)$ where $S$ is a finite set of states and $\probmat : S \rightarrow \distrset(S)$ is a total stochastic transition function. In the sequel, $\probmat$ is sometimes seen as a matrix, and for all $s, s' \in S$, we write $\probmat(s, s')$ for $\probmat(s)(s')$. 
A \textit{path} is an infinite sequence of states $\rho = s_0s_1s_2 \ldots $ such that $\probmat(s_i, s_{i+1}) > 0$ for all $i \geq 0$. Finite paths are defined similarly, and $\probmat$ is naturally extended to finite paths.

A \textit{Markov decision process (MDP)} is a tuple $(S, \ActionsO, \probmat)$ where $S$ is a finite set of states, $\ActionsO$ is a finite set of actions, and $\probmat : S \times \ActionsO \rightarrow \distrset(S)$ is a partial stochastic transition function. We often write $\probmat(s, \actionO, s')$ for $\probmat(s, \actionO)(s')$. For each $s \in S$, we denote by $\enabledactions \subseteq \ActionsO$ the set of enabled actions in $s$, where an action $\actionO \in \ActionsO$ is \textit{enabled} in $s$ if $(s, \actionO) \in \domain(\probmat)$. We require  $\forall s \in S$, $\enabledactions \neq \emptyset$, that is, the MDP is $\ActionsO$\textit{-non-blocking}. For each $\actionO \in \ActionsO$, $S_\actionO$ denotes the set of states in which $\actionO$ is enabled.

Let $M = (S, \ActionsO, \probmat)$ be an MDP. A \textit{memoryless strategy} is a total function $\lambda : S \rightarrow \ActionsO$ mapping each state $s$ to an enabled action $\actionO \in \enabledactions$. We denote by $\Lambda$ the set of all memoryless strategies. A memoryless strategy $\lambda$ induces an MC $M_{\lambda} = (S, \probmat_\lambda)$ such that for all $s, s' \in S$, $\probmat_\lambda(s, s') = \probmat(s, \lambda(s), s')$.

\paragraph{Costs and value functions.}
In addition to an MDP $M = (S, \ActionsO, \probmat)$, we consider a partial \textit{cost function} $\reward : S \times \ActionsO \rightarrow \R$ with $\domain(\reward) = \domain(\probmat)$ that associates a cost with $s \in S$ and $\actionO \in \enabledactions$. A memoryless strategy $\lambda$ assigns a total cost function $\reward_\lambda : S \rightarrow \R$ to the induced MC $M_{\lambda}$, such that $\reward_\lambda(s) = \reward(s, \lambda(s))$. 
Given a path $\rho = s_0s_1s_2 \ldots $ in $M_{\lambda}$, 
the \textit{mean-payoff} of $\rho$ is $\MP(\rho) = \limsup_{n \rightarrow \infty} \frac{1}{n}\sum_{i = 0}^{n-1}  \reward_\lambda(s_i)$. Given a subset $G \subseteq S$ of \textit{goal} states and a finite path $\rho$ reaching a state of $G$, the \textit{truncated sum up to $G$} of $\rho$ is $\TS_G(\rho) = \sum_{i = 0}^{n-1}  \reward_\lambda(s_i)$ where $n$ is the first index such that $s_n \in G$.

Given an MDP with a cost function $\reward$, and a memoryless strategy $\lambda$, we consider two classical value functions of $\lambda$. 
For all states $s \in S$, 
the \textit{expected mean-payoff} of $\lambda$ is $\EMP_\lambda(s) = \lim_{n \to \infty} \frac{1}{n} \sum_{i=0}^{n-1}\probmat_\lambda^i \reward_\lambda (s)$. Given a subset $G \subseteq S$, and assuming that $\lambda$ reaches $G$ from state $s$ with probability 1, the \textit{expected truncated sum up to $G$} of $\lambda$ is $\ETP_\lambda(s) = \sum_{\rho}\probmat_\lambda(\rho) \TS_G (\rho)$ where the sum is over all finite paths $\rho = s_0 s_1 \ldots s_n$ such that $s_0 = s$, $s_n \in G$, and $s_0, \ldots, s_{n-1} \not \in G$.
Let $\lambda^*$ be a memoryless strategy. Given a value function $\Edot_{\lambda} \in \{
\EMP_{\lambda},\ETP_{\lambda}\}$, we say that $\lambda^*$ is \textit{optimal} if $\Edot_{\lambda^*}(s) = \inf_{\lambda \in \Lambda} \Edot_{\lambda}(s)$ for all $s \in S$, and $\Edot_{\lambda^*}$ is called the \textit{optimal} value function.\footnote{\label{fn:altobj}An alternative objective is to maximize the value function, in which case $\lambda^*$ is optimal if $\Edot_{\lambda^*}(s) = \sup_{\lambda \in \Lambda} \Edot_{\lambda}(s)$, $\forall s \in S$.} 
Note that we might have considered other classes of strategies, but for these value functions, there always exists a memoryless strategy minimizing the expected value of all states~\cite{DBLP:books/daglib/0020348,putermanMDP}. 

\begin{figure}[h]
\centering
\vspace*{-0.2cm}
  \begin{tikzpicture}[>=stealth',shorten >=1pt,auto,node distance=2.5cm,bend angle=0,scale=1,font=\small]
   \tikzstyle{p1}=[draw,circle,text centered,minimum size=2.5mm]
    \node[p1]  (0)  at (0, 0) {$s_0$};
    \node[p1]  (1)  at (2.5, 0.6) {$s_1$};
    \node[p1]  (2)  at (2.5, -0.6) {$s_2$};
     \fill (1.12,0) circle (0.04cm);

    \node[p1]  (3)  at (7.6, 0) {$s_0$};
    \node[p1]  (4)  at (10.1, 0.6) {$s_1$};
    \node[p1]  (5)  at (10.1, -0.6) {$s_2$};
     \fill (8.72,0) circle (0.04cm);

    \node at (4.7,0.5) {$\probmat(s_0, \sigma, s_0) = \frac{1}{2}$};
    \node at (4.7,0) {$\probmat(s_0, \sigma, s_1) = \frac{1}{6}$};     
    \node at (4.7,-0.5) {$\probmat(s_0, \sigma, s_2) = \frac{1}{3}$};

    \node at (12.4,1.4) {$\edges(s_0, \sigma)(\actionI_0) = s_0$};
    \node at (12.4,0.9) {$\edges(s_0, \sigma)(\actionI_1) = s_1$};     
    \node at (12.4,0.4) {$\edges(s_0, \sigma)(\actionI_2) = s_2$};
    \node at (12.4,-0.2) {$\distr(s_0, \sigma)(\actionI_0) = \frac{1}{2}$};
    \node at (12.4,-0.7) {$\distr(s_0, \sigma)(\actionI_1) = \frac{1}{6}$};     
    \node at (12.4,-1.2) {$\distr(s_0, \sigma)(\actionI_2) = \frac{1}{3}$};

    \draw (6.5,-1.25) --  (6.5,1.75);

    \node at (0.4,1.5) {$M = (S, \ActionsO, \probmat)$};
    \node at (8.5,1.5) {$M = (S, \ActionsO, \ActionsI, \edges, \distr)$};
    \path
	(0) -- (0);
    \draw[->,>=latex] (0) to[out=0,in=180] node [above, xshift=0mm,yshift=0mm] {$\actionO$} (1.1, 0);
    \draw[->,>=latex] (1.12, -0.06) to[out=270,in=270, distance=0.8cm] node [above, xshift=1mm,yshift=0mm] {$\frac{1}{2}$} (0);
    \draw[->,>=latex] (1.16, 0.05) to[out=60,in=180] node [above, xshift=0mm] {$\frac{1}{6}$} (1);
    \draw[->,>=latex] (1.16, -0.05) to[out=300,in=180] node [above, xshift=0mm,yshift=0mm] {$\frac{1}{3}$} (2);

    \draw[->,>=latex] (3) to[out=0,in=180] node [above, xshift=0mm,yshift=0mm] {$\actionO$} (8.7, 0);
    \draw[->,>=latex] (8.72, -0.06) to[out=270,in=270, distance=0.8cm] node [above, xshift=0.5mm,yshift=0mm] {$\actionI_0$} (3);
    \draw[->,>=latex] (8.76, 0.05) to[out=60,in=180] node [above, xshift=0mm] {$\actionI_1$} (4);
    \draw[->,>=latex] (8.76, -0.05) to[out=300,in=180] node [above, xshift=0mm,yshift=0mm] {$\actionI_2$} (5);
\end{tikzpicture}
\vspace{-0.5cm}
\caption{Illustration of the new definition of MDPs for a state $s_0 \in S$ and an action $\actionO \in \Sigma_{s_0}$.}
\label{fig:exmdp}
\end{figure}
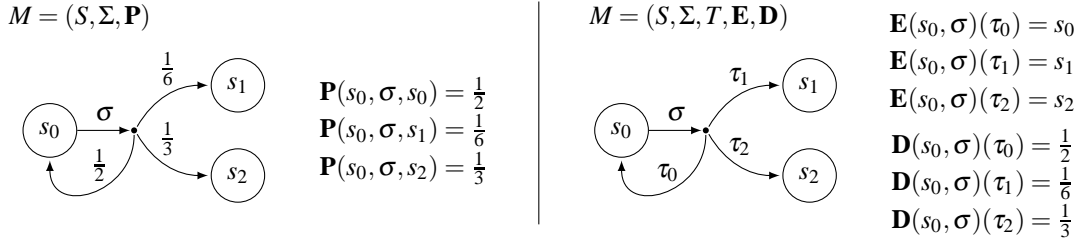

\paragraph{Studied problems.} In this paper, we study algorithms for solving MDPs for two quantitative settings: the expected mean-payoff and the stochastic shortest path. Let $M$ be an MDP with a cost function $\reward$. $(i)$~The \textit{expected mean-payoff (EMP) problem} is to synthesize an optimal strategy for the expected mean-payoff value function.
$(ii)$~When $\reward$ is restricted to 
\textit{strictly positive} values in $\R_{> 0}$, and a subset $G \subseteq S$ of goal states is given, the \textit{stochastic shortest path (SSP) problem} is to synthesize  an optimal strategy for the expected truncated sum value function, among the set of strategies that reach $G$ with probability $1$, provided such strategies exist. For all $s \in S$, we denote by $\Lambda_s^P$ the set of \textit{proper strategies} that lead from $s$ to $G$ with probability $1$. Solving the SSP problem consists of two steps. The first step is to determine the set $S^P = \{s \in S \mid \Lambda_s^P \neq \emptyset\}$ of \textit{proper states} that have at least one proper strategy. The second step consists in synthesizing an optimal strategy $\lambda^*$ such that $\ETP_{\lambda^*}(s) = \inf_{\lambda \in \Lambda_s^P}\ETP_\lambda(s)$ for all  $s \in S^P$. It is known that both problems can be solved in polynomial time via linear programming, with memoryless optimal strategies~\cite{putermanMDP,filar1996competitive,bertsekas1996neuro,bertsekas1991analysis}. 
\paragraph{Monotonic MDPs.} 
In this paper, we need a slightly different, but equivalent, definition of MDPs based on the next idea. Instead of a transition function $\probmat : S \times \ActionsO \rightarrow \distrset(S)$, we rather use two functions in a way to separate the probabilities from the successors as indicated in Figure~\ref{fig:exmdp}. In this new definition, an MDP is a tuple $M = (S, \ActionsO, \ActionsI, \edges, \distr)$ where $S$ is a finite set of states, $\ActionsO$ and $\ActionsI$ are two disjoint finite sets of actions, $\edges : S \times \ActionsO \rightarrow \totfunc(\ActionsI, S)$ is a partial successor function, and $\distr : S \times \ActionsO \rightarrow \distrset(\ActionsI)$ is a partial stochastic function such that $\domain(\edges) = \domain(\distr)$. In this context, the notion of MC $M_{\lambda}$ induced by a strategy $\lambda$ is defined as $(S, \ActionsI, \edges_{\lambda}, \distr_{\lambda})$ with functions $\edges_{\lambda}, \distr_{\lambda}$ naturally defined. In the sequel, depending on the context, we will use both definitions.

Let $S$ be a finite set equipped with a partial order $\preceq$ such that $(S,\preceq)$ is a \textit{semilattice}, i.e. for all $s, s' \in S$, their greatest lower bound $s \sqcap s'$ always exists. A set $L \subseteq S$ is \textit{closed} for $\preceq$ if for all $s \in L$ and all $s' \preceq s$, we have $s' \in L$. 

A \textit{monotonic MDP} is an MDP $\monMDP = (S, \ActionsO, \ActionsI, \edges, \distr)$ such that:
\begin{enumerate} 
	\item $S$ is equipped with a partial order $\preceq$ such that $(S, \preceq)$ is a semilattice, and
	\item $\preceq$ is \textit{compatible} with $\edges$, i.e. for all $s, s' \in S$, if $s \preceq s'$, then  for all $\actionO \in \ActionsO$, $\actionI \in \ActionsI$, for all $t' \in S$ such that $\edges(s', \actionO)(\actionI) = t'$, there exists $t \in S$ such that $\edges(s, \actionO)(\actionI) = t$ and $t \preceq t'$.
\end{enumerate}


\noindent
Note that since $(S, \preceq)$ is a semilattice, $S$ is trivially closed for $\preceq$. With this definition, we have the next important properties: (1) for all $s,s' \in S$, if $s \preceq s'$ then $\enabledactionsprime \subseteq \enabledactions$, and (2) for all $\actionO \in \ActionsO$, $\enabledstates$ is closed. 

\begin{remark}
In this definition, by monotonic MDPs, we mean MDPs that are built on state spaces \textit{already equipped with a natural partial order}. For instance, this is the case for the two classes of MDPs studied in Section~\ref{sec:experiments}. 
The same kind of approach has already been proposed in~\cite{DBLP:journals/tcs/FinkelS01}.

Note that all MDPs can be seen monotonic. Indeed, let $(S, \Sigma, T, \edges, \distr)$ be a given MDP and let $\preceq$ be a partial order such that all states in $S$ are pairwise incomparable with respect to $\preceq$. By adding a new state~$t$ such that $t \preceq s$, for all $s \in S$, and such that $t$ is an isolated state with a self-loop, we have that $(S\cup\{t\}, \preceq)$ is a semilattice and $\preceq$ is compatible with $\edges$. However, such a partial order would not lead to efficient algorithms in the sense studied in this paper. 
\end{remark}


\section{Strategy iteration algorithms and symblicit approach}\label{sec:si}
In this section, we first present strategy iteration algorithms for synthesizing optimal strategies for both the SSP and EMP problems. We then present a symblicit version of those algorithms that mixes symbolic and explicit data representations.
Our presentation  is inspired from the one given in \cite{DBLP:conf/qest/WimmerBBHCHDT10} for the EMP problem.

%

\begin{figure*}[t]
\vspace{-0.5cm}
 \begin{tabular}{lcr}
 \begin{minipage}[t]{.44\textwidth}
\begin{algorithm}[H]
\caption{\footnotesize{\textsc{SSP\_SI}$($MDP $M$, Strictly positive cost function $\reward$, Goal states $G)$}}
{\scriptsize
\begin{algorithmic}[1] \label{algo:SSPSI}
\STATE $n := 0, \lambda_n :=$ \textsc{InitialProperStrat}$(M, G)$
\REPEAT
	\STATE Obtain $v_n$ by solving 
	\vspace*{-0.1cm}
	\begin{center}
	$\reward_{\lambda_n} + (\probmat_{\lambda_n} - \id)v_n = 0$
	\end{center}
 	\STATE \begin{flushleft}$\widehat{\enabledactions} := \underset{\actionO \in \ActionsO_s}{\arg \min} (\reward(s,\actionO) + \underset{s'\in S}{\sum}\probmat(s, \actionO, s')\cdot v_n(s')), \forall s \in S$\end{flushleft}
	\STATE Choose $\lambda_{n+1}$ s.t. $\lambda_{n+1}(s) \in \widehat{\enabledactions}, \forall s \in S$, \\
	\hspace{0.5cm} setting $\lambda_{n+1}(s) := \lambda_n(s)$ if possible.
	\STATE $n := n+1$
\UNTIL{$\lambda_{n} = \lambda_{n-1}$}
\STATE return $(\lambda_{n-1}, v_{n-1})$
\end{algorithmic}
}
\end{algorithm}
 \end{minipage}
 &
&
 \begin{minipage}[t]{.48\textwidth}

\begin{algorithm}[H]
\caption{{\footnotesize\textsc{Symblicit}(MDP $\symbMDP$, [Strictly positive] cost function $\symbReward$[, Goal states $\symbGoal$])}}
{\scriptsize
\begin{algorithmic}[1] \label{algo:symblicit}
\STATE $n := 0, \lambda_n :=$ \textsc{InitialStrat}$(\symbMDP[, \symbGoal])$
\REPEAT
	\STATE $(\symbMC, \symbRewardMC) :=$ \textsc{InducedMC\&Cost}$(\symbMDP, \symbReward, \lambda_n)$
	\STATE $(\symbQuotient, \symbRewardQuotient) :=$ \textsc{Lump}$(\symbMC, \symbRewardMC)$
	\STATE $(\explQuotient, \explRewardQuotient) :=$ \textsc{Explicit}$(\symbQuotient, \symbRewardQuotient)$
	\STATE $x_n :=$ \textsc{SolveLinearSystem}$(\explQuotient, \explRewardQuotient)$			
	\STATE $\symbValue_n :=$ \textsc{Symbolic}$(x_n)$
	\STATE $\lambda_{n+1} :=$ \textsc{ImproveStrat}$(\symbMDP, \lambda_n, \symbValue_n)$	
	\STATE $n := n+1$
\UNTIL{$\lambda_{n} = \lambda_{n-1}$}
\STATE return $(\lambda_{n-1}, \symbValue_{n-1})$
\end{algorithmic}
}
\end{algorithm}
 \end{minipage}
 \end{tabular}
\end{figure*}

\paragraph{Algorithm for the SSP problem.}
Let $M = (S, \ActionsO, \probmat)$ be an MDP, $\reward : S \times \ActionsO \rightarrow \R_{> 0}$ be a strictly positive cost function, and $G \subseteq S$ be a set of goal states. As explained in Section~\ref{sec:prelim}, the set $S^P$ of proper states is computed in a preliminary step (see~\cite{DBLP:conf/concur/Alfaro99} for a quadratic algorithm). The strategy iteration algorithm~\cite{howard1960dynamic,bertsekas1996neuro} (see Algorithm~\ref{algo:SSPSI}) is then applied under the typical assumption that all cycles in the underlying graph of $M$ have strictly positive cost~\cite{bertsekas1996neuro}. This assumption holds in our case by definition of the cost function~$\reward$. 
Algorithm~\ref{algo:SSPSI} starts with an arbitrary proper strategy $\lambda_0$, that can be easily computed with the algorithm of~\cite{DBLP:conf/concur/Alfaro99}, and improves it until an optimal strategy is found. The expected truncated sum $v_n$ up to $G$ of the current strategy $\lambda_n$ is computed by solving the system of linear equations in line 3, and used to improve the strategy (if possible) at each state. Note that the strategy $\lambda_n$ is improved at a state $s$ to an action $\sigma \in \enabledactions$ only if the new expected truncated sum is strictly smaller than the expected truncated sum of the action $\lambda_n(s)$, i.e. only if $\lambda_n(s) \not\in \underset{\sigma \in \enabledactions}{\arg\min}(\reward(s,\actionO) + \underset{s'\in S}{\sum}\probmat(s, \actionO, s')\cdot v_n(s'))$. If no improvement is possible for any state, an optimal strategy is found and the algorithm terminates in line 7. Otherwise, it restarts by solving the new equation system, tries to improve the strategy using the new values computed, and so~on.


\paragraph{Algorithm for the EMP problem.}
The strategy iteration algorithm for the EMP problem works similarly (see~\cite{veinott1966finding,putermanMDP} for more details). The algorithm starts with an arbitrary strategy $\lambda_0$. Solving the related linear system leads to two values: the gain value $g_n$ and bias value $b_n$ of strategy $\lambda_n$. The gain corresponds to the expected mean-payoff, while the bias can be interpreted as the expected total difference between the cost and the expected mean-payoff. The computed gain value is used to locally improve the strategy. If such an improvement is not possible for any state, the bias value is used to locally improve the strategy such that only actions that also optimize the gain are considered. 

\paragraph{Bisimulation lumping.}
When treating MDPs and induced MCs with large state spaces, using explicit representations quickly becomes unfeasible for the algorithms presented above. Given an MC $(S, \probmat)$  and a cost function $\reward : S \rightarrow \R$, the \textit{bisimulation lumping} technique~\cite{kemeny1960jl,DBLP:journals/iandc/LarsenS91,buchholz1994exact} consists in gathering certain states of $S$ which behave equivalently according to the class of properties under consideration. Let $\sim$ be an equivalence relation on $S$ and $S_{\sim}$ be the induced \textit{partition}. We call \textit{block} of $S_{\sim}$ any equivalence class of $\sim$.
We say that $\sim$ is a \textit{bisimulation} if for all $s,t \in S$ such that $s \sim t$, we have $\reward(s) = \reward(t)$ and $\probmat(s,C) = \probmat(t,C)$ for all block $C \in S_{\sim}$, where $\probmat(s,C) = \sum_{s'\in C}\probmat(s,s')$. 
%
%
The \textit{bisimulation quotient} is the MC $(S_{\sim}, \probmat_{\sim})$ such that $\probmat_{\sim}(C,C') = \probmat(s,C')$, where $s \in C$ and $C,C' \in S_{\sim}$. The cost function $\reward_{\sim} : S_{\sim} \rightarrow \R$ is transferred to the quotient such that $\reward_{\sim}(C) = \reward(s)$, where $s \in C$ and $C \in S_{\sim}$. This quotient is a minimized model equivalent to the original, since it keeps the expected truncated sum and expected mean-payoff as in the original model~\cite{DBLP:journals/iandc/BaierKHW05}. Usually, we are interested in computing the unique \textit{largest} bisimulation, denoted $\equivlump$, which leads to the smallest bisimulation quotient $(S_{\equivlump}, \probmat_{\equivlump})$ (see~\cite{DBLP:journals/ipl/DerisaviHS03} for an algorithm).

\paragraph{Symblicit algorithm.} 

The symblicit algorithm is described in Algorithm~\textsc{Symblicit} for both the SSP and EMP (see Algorithm~\ref{algo:symblicit}). 
It combines symbolic\footnote{We use caligraphic style for symbols denoting a symbolic representation.} and explicit representations of the manipulated data as follows. The MDP $\symbMDP$, the cost function $\symbReward$, the strategies $\lambda_n$, the induced MCs $\symbMC$ with cost functions $\symbRewardMC$, and the set $\symbGoal$ of goal states for the SSP, are symbolically represented. The lumping algorithm is applied on symbolic MCs and produces a symbolic representation of the bisimulation quotient $\symbQuotient$ with cost function $\symbRewardQuotient$ (line 4). 
The computed quotient is converted to a sparse matrix representation (line~5). As it is in general much smaller than the original model, there is no memory issue by storing it explicitly, and the linear system can be solved. The computed value functions $x_n$ (corresponding to $v_n$ for the SPP, and $g_n$ and $b_n$ for the EMP) are then converted into symbolic representations $\symbValue_n$, and transferred back to the original MDP (line 7). Finally, the update of the strategy is performed symbolically.

\section{Pseudo-antichains}\label{sec:pa}

We want to develop a symblicit algorithm for solving the SSP and EMP problems for monotonic MDPs. To this end, we here introduce a new data structure extended from antichains, called pseudo-antichains.

\paragraph{Closed sets and antichains.} Let $(S,\preceq)$ be a semilattice.
The \textit{closure} $\antclos L$ of a set $L \subseteq S$ is the set $\antclos L = \{s' \in S$ $|$ $\exists s \in L \cdot s' \preceq s\}$. 
A set $\alpha$ is an \textit{antichain} if all its elements are pairwise incomparable with respect to $\preceq$. For $L \subseteq S$, we denote by $\lceil L \rceil$ the antichain of its maximal elements.
If $L$ is closed, then  $\antclos\lceil L \rceil = L$, and $\lceil L \rceil$ is called the \textit{canonical representation} of $L$. The interest of antichains is that they are \textit{compact representations of closed sets}. Antichain based algorithms have proved worthy for solving classical problems in game theory, but also in logic and automata theory (e.g. \cite{DBLP:conf/cav/WulfDHR06,DBLP:conf/tacas/DoyenR07,DBLP:conf/wia/BouajjaniHHTV08,DBLP:conf/tacas/DoyenR10}).
We have the following classical properties on antichains (see for instance \cite{DBLP:journals/fmsd/FiliotJR11}):

\begin{proposition} \label{prop:opantichains}
Let $\alpha_1, \alpha_2 \subseteq S$ be two antichains and $s \in S$. Then:
\begin{itemize}
\item $s \in\  \antclos \alpha_1$ iff $\exists a \in \alpha_1 \cdot s \preceq a$
\item  $\antclos\alpha_1$ $\cup$ $\antclos\alpha_2 =\ \antclos\lceil \alpha_1 \cup \alpha_2\rceil$
\item  $\antclos\alpha_1$ $\cap$ $\antclos\alpha_2 =\ \antclos\lceil \alpha_1 \sqcap \alpha_2\rceil$, 
where $\alpha_1 \sqcap \alpha_2 \defeq \{a_1 \sqcap a_2 \mid a_1 \in \alpha_1,  a_2 \in \alpha_2 \}$
\item $\antclos\alpha_1 \subseteq\ \antclos\alpha_2$ iff $\forall a_1 \in \alpha_1 \cdot \exists a_2 \in \alpha_2 \cdot a_1 \preceq a_2$
\end{itemize}
\end{proposition}

\noindent
For convenience, when $\alpha_1$ and $\alpha_2$ are antichains, we use notation $\alpha_1$ $\antunion$ $\alpha_2$ (resp. $\alpha_1$ $\antinter$ $\alpha_2$) for the antichain $\lceil\antclos\alpha_1$ $\cup \antclos\alpha_2\rceil$ (resp. $\lceil\antclos\alpha_1$ $\cap \antclos\alpha_2\rceil$). Let $L_1, L_2 \subseteq S$ be two closed sets. Unlike the union or intersection, the difference $L_1\diff L_2$ is not necessarily a closed set. There is thus a need for a new structure that ``represents" $L_1 \diff L_2$ in a compact way.

\paragraph{Pseudo-elements and pseudo-closures.} 
A \textit{pseudo-element} is a pair $(x, \alpha)$ where $x \in S$ and $\alpha \subseteq S$ is an antichain such that $x\not\in$ $\antclos\alpha$. The \textit{pseudo-closure} of a pseudo-element $(x, \alpha)$, denoted by $\pseudclos(x, \alpha)$, is the set $\pseudclos(x, \alpha) = \{s \in S \mid s \preceq x$ and $s \not\in$ $\antclos\alpha\} =$ $\antclos\{x\}\diff\!\!\antclos\alpha$. Notice that $\pseudclos(x, \alpha)$ is non empty since $x\not\in$ $\antclos\alpha$ by definition of a pseudo-element. The following example illustrates these notions.
\begin{example} \label{ex:pseudo-element}
Let $\nat^2_{\leq3}$ be the set of pairs of natural numbers in $[0, 3]$ and let $\preceq$ be a partial order on $\nat^2_{\leq3}$ such that $(n_1, n_1') \preceq (n_2, n_2')$ iff $n_1 \leq n_2$ and $n_1' \leq n_2'$. Then, $(\nat^2_{\leq3}, \preceq)$ is a complete lattice with least upper bound $\sqcup$ such that $(n_1, n_1') \sqcup (n_2, n_2') = (\max(n_1, n_2), \max(n_1', n_2'))$, and greatest lower bound $\sqcap$ such that $(n_1, n_1') \sqcap (n_2, n_2') = (\min(n_1, n_2), \min(n_1', n_2'))$.
With $x = (3, 2)$ and $\alpha = \{(2, 1), (0, 2)\}$, the pseudo-closure of the pseudo-element $(x, \alpha)$ is the set $\pseudclos(x, \alpha) = \{(3, 2), (3, 1), (3, 0), (2, 2), (1, 2)\} =\ \antclos\{x\}\setminus\!\!\antclos\alpha$ (see Figure~\ref{fig:pa}).
\end{example}

\begin{figure}
\centering
\begin{minipage}{0.47\linewidth}
     \centering
          \begin{tikzpicture}[-,>=stealth',auto,node distance=2.5cm,bend angle=45,scale=0.68,font=\footnotesize]
	     \fill[light-gray] (-3.7,3) --  (-1, 5.7) -- (2.7, 2) -- (2, 2.7) -- (1, 1.7) -- (-1, 3.7) -- (-2.7, 2);
	    \node (0)  at (0, 0) {$(0,0)$};
	    \node (1)  at (-1,1) {$(1, 0)$};
	    \node (2)  at (1,1) {$(0, 1)$};
	    \node (3)  at (0,2) {$(1, 1)$};
	    \node (4)  at (-2,2) {$(2, 0)$};
	    \node (5)  at (2,2) {$(0, 2)$};
	    \node (6)  at (-3, 3) {$(3, 0)$};
	    \node (7)  at (-1,3) {$(2, 1)$};
	    \node (8)  at (1,3) {$(1, 2)$};
	    \node (9)  at (3,3) {$(0, 3)$};
	    \node (10)  at (-2,4) {$(3, 1)$};
	    \node (11)  at (0,4) {$(2, 2)$};
	    \node (12)  at (2,4) {$(1, 3)$};
	    \node (13)  at (-1,5) {$(3, 2)$};
	    \node (14)  at (1,5) {$(2, 3)$};
	    \node (15)  at (0,6) {$(3, 3)$};
	   \path
		(0) edge[dashed] (1)
	         (0) edge[dashed] (2)
		(1) edge[dashed] (3)
		(1) edge[dashed] (4)
		(2) edge[dashed] (3)
		(2) edge[dashed] (5)
		(3) edge[dashed] (7)
		(3) edge[dashed] (8)
		(4) edge[dashed] (6)
		(4) edge[dashed] (7)
		(5) edge[dashed] (8)
		(5) edge[dashed] (9)
		(6) edge[dashed] (10)
		(7) edge[dashed] (10)
		(7) edge[dashed] (11)
		(8) edge[dashed] (11)
		(8) edge[dashed] (12)
		(9) edge[dashed] (12)
		(10) edge[dashed] (13)
		(11) edge[dashed] (13)
		(11) edge[dashed] (14)
		(12) edge[dashed] (14)
		(13) edge[dashed] (15)
		(14) edge[dashed] (15)
	
		(-3.7,3) edge (-1, 5.7)
		(2.7,2) edge (-1, 5.7)
	
		(-2.7, 2) edge (-1, 3.7)
		(1, 1.7) edge (-1, 3.7)
		(2, 2.7) edge (1, 1.7)
		(2.7, 2) edge (2, 2.7);
	\end{tikzpicture}
	\vspace{-0.36cm}
	\caption{Pseudo-closure of a pseudo-element over $(\nat^2_{\leq3}, \preceq)$.}
	\label{fig:pa}
   \end{minipage}
   \hfill
   \begin{minipage}{0.47\linewidth}
	\centering
               \begin{tikzpicture}[-,>=stealth',auto,node distance=2.5cm,bend angle=45,scale=0.83,font=\footnotesize]
		     \fill[light-gray2] (-2.5,-0.5) -- (-0.5,1.5) -- (2,-1) -- (1, -2) -- (0.5,-1.5) -- (0,-2) -- (-1,-1) -- (-1.5,-1.5);
		     \fill[light-gray] (-2,-1) -- (-0.5,0.5) -- (1.5,-1.5) -- (1, -2) -- (-0.5,-0.5) -- (-1.5,-1.5);
		    \node (0)  at (-0.5, 1.8) {$y$};
		    \node (0)  at (0.5, -1.2) {$b_2$};
		    \node (0)  at (-0.5, 0.8) {$x$};
		    \node (0)  at (-1, -0.7) {$b_1$};
		    \node (0)  at (-0.5, -0.2) {$a$};
		    \node (0)  at (-0.5, 1.5) {$\bullet$};
		    \node (0)  at (-0.5, 0.5) {$\bullet$};
		    \node (0)  at (0.5, -1.5) {$\bullet$};    
		    \node (0)  at (-1, -1) {$\bullet$};    
		    \node (0)  at (-0.5, -0.5) {$\bullet$};
		
		    \path
			(-0.5,1.5) edge (-2.5,-0.5)
			(-0.5,1.5) edge (2,-1)
			(-0.5,0.5) edge (-2,-1)
			(-0.5,-0.5) edge (-1.5,-1.5)
			(-0.5,0.5) edge (1.5,-1.5)
			(-0.5,-0.5) edge (1,-2)
			(-1,-1) edge (0,-2)
			(0.5,-1.5) edge (0,-2)
			
			(-2.5,-0.5) edge[ultra thin,dashed] (0.5,2.5)
			(-2,-1) edge[ultra thin,dashed] (1,2)
			(-1.5,-1.5) edge[ultra thin,dashed] (1.5,1.5)
			(-1,-2) edge[ultra thin,dashed] (2,1)
			(-0.5,-2.5) edge[ultra thin,dashed] (2.5,0.5)
			(0,-3) edge[ultra thin,dashed] (3, 0)
			(0,-3) edge[ultra thin,dashed] (-3,0)
			(0.5,-2.5) edge[ultra thin,dashed] (-2.5,0.5)
			(1,-2) edge[ultra thin,dashed] (-2,1)
			(1.5,-1.5) edge[ultra thin,dashed] (-1.5,1.5)
			(2,-1) edge[ultra thin,dashed] (-1,2)
			(2.5,-0.5) edge[ultra thin,dashed] (-0.5,2.5);
	\end{tikzpicture}
		\vspace{-0.36cm}
	\caption[Inclusion of pseudo-closures of pseudo-elements]{Inclusion of pseudo-closures of pseudo-elements.}
	\label{fig:pa_inclusion}
     \end{minipage}
\vspace*{-0.45cm}
\end{figure}

There may exist two pseudo-elements $(x, \alpha)$ and $(y, \beta)$ such that  $\pseudclos(x, \alpha) =$  $\pseudclos(y, \beta)$. A pseudo-element $(x, \alpha)$ is said to be in \textit{canonical form} if $\forall a \in \alpha \cdot a \preceq x$ (as in Example~\ref{ex:pseudo-element}). 
Notice that for all pseudo-elements $(x, \alpha)$, there exists a pseudo-element in canonical form $(y, \beta)$ such that $\pseudclos(x, \alpha) =$ $\pseudclos(y, \beta)$: it is equal to $(x, \{x\}\ \antinter\ \alpha)$ and called the \textit{canonical representation} of $\pseudclos(x, \alpha)$.
The next corollary of Proposition~\ref{prop:inclusion} shows that the canonical form is unique. 

\begin{proposition} \label{prop:inclusion}
Let $(x, \alpha)$ and $(y, \beta)$ be two pseudo-elements. Then $\pseudclos(x, \alpha) \subseteq$ $\pseudclos(y, \beta)$ iff $x \preceq y$ and $\forall b \in \beta \cdot b \sqcap x \in$ $\antclos\alpha$.
\end{proposition}
%

\begin{corollary}
Let $(x, \alpha)$ and $(y, \beta)$ be two pseudo-elements in canonical form. Then $\pseudclos(x, \alpha)$ $=$ $\pseudclos(y, \beta)$ iff $x = y$ and $\alpha = \beta$.
\end{corollary}
%
%

The following example illustrates Proposition~\ref{prop:inclusion}. 

\begin{example}
Let $(S, \preceq)$ be a semilattice and let $(x, \{a\})$ and $(y, \{b_1,b_2\})$, with $x, y, a, b_1, b_2 \in S$, be two pseudo-elements as depicted in Figure~\ref{fig:pa_inclusion}. The pseudo-closure of $(x, \{a\})$ is depicted in dark gray, whereas the pseudo-closure of $(y, \{b_1,b_2\})$ is depicted in (light and dark) gray. We have $x \preceq y$, $b_1\sqcap x = b_1 \in$ $\antclos\{a\}$ and $b_2\sqcap x = b_2 \in$ $\antclos\{a\}$. Therefore $\pseudclos(x, \{a\}) \subseteq$ $\pseudclos(y, \{b_1,b_2\})$.
\end{example}

\paragraph{Pseudo-antichains.}
A \textit{pseudo-antichain} $A$ is a finite set of pseudo-elements, that is $A = \{(x_i, \alpha_i) \ | \ i \in I\}$ with $I$ finite.
The \textit{pseudo-closure} $\pseudclos A$ of $A$ is defined as the set $\pseudclos A = \bigcup_{i\in I}\pseudclos(x_i, \alpha_i)$. Given $(x_i, \alpha_i)$, $(x_j, \alpha_j) \in A$, 
we observe that: (1) if $x_i = x_j$, then $(x_i, \alpha_i)$ and $(x_j, \alpha_j)$ can be replaced in $A$ by the pseudo-element $(x_i, \alpha_i\ \antinter\ \alpha_j)$, and (2) if $\pseudclos(x_i, \alpha_i) \subseteq$ $\pseudclos(x_j, \alpha_j)$, then $(x_i, \alpha_i)$ can be removed from $A$.
Therefore, we say that a pseudo-antichain $A = \{(x_i, \alpha_i) \ | \ i \in I\}$ is \textit{simplified} if $\forall i \cdot (x_i, \alpha_i)$ is in canonical form, and $\forall i \neq j \cdot x_i \neq x_j$ and $\pseudclos(x_i, \alpha_i) \not\subseteq$ $\pseudclos(x_j, \alpha_j)$. 
Notice that two distinct pseudo-antichains $A$ and $B$ can have the same pseudo-closure $\pseudclos A =$ $\pseudclos B$ even if they are simplified. We thus say that $A$ is a \textit{\PA-representation}\footnote{``\PA-representation" means pseudo-antichain based representation.} of $\pseudclos A$ (without saying that it is a canonical representation), and that $\pseudclos A$ is \textit{\PA-represented} by $A$. 

Any antichain $\alpha$ can be seen as the pseudo-antichain $A = \{(x, \emptyset) \mid x \in \alpha\}$. Furthermore, notice that \textit{any set} $X$ is \PA-represented by $A = \{(x, \alpha_x) \ | \ x \in X\}$, with  $\alpha_x = \lceil \{s \in S \ | \ s \preceq x \textnormal{ and } s \neq x\}\rceil$. Indeed $\pseudclos (x, \alpha_x) =  \{x\}$ for all $x$, and thus $X =$ $\pseudclos A$. 

The interest of pseudo-antichains is that they behave well with respect to \textit{all Boolean operations}, as shown by  Proposition \ref{prop:BooleanOp} on pseudo-elements (and easily extended to pseudo-antichains). From the algorithmic point of view, it is important to note that the computations only manipulate (pseudo-)antichains instead of their (pseudo-)closure. 
Note also that the pseudo-antichains computed in this proposition are not necessarily simplified. However, our algorithms implementing those operations always simplify the computed pseudo-antichains for the sake of efficiency.

\begin{proposition} 
\label{prop:BooleanOp}
Let $(x, \alpha), (y, \beta)$ be two pseudo-elements.
Then:
\begin{itemize}
\item $\pseudclos(x, \alpha)$ $\cup$ $\pseudclos(y, \beta) =$ $\pseudclos\{(x, \alpha), (y, \beta)\}$
\item $\pseudclos(x, \alpha)$ $\cap$ $\pseudclos(y, \beta) =$ $\pseudclos\{(x\sqcap y, \alpha$ $\antunion$ $\beta)\}$
\item $\pseudclos(x, \alpha)$ $\diff$ $\pseudclos(y, \beta) =$ $\pseudclos \big( \{(x, \{y\}$ $\antunion$ $\alpha) \} \cup \{ (x\sqcap b, \alpha) \mid b \in \beta \} \big)$
\end{itemize}
\end{proposition}

\begin{figure}[t]
\centering
  \begin{tikzpicture}[-,>=stealth',auto,node distance=2.5cm,bend angle=45,scale=0.85,font=\footnotesize]
    \fill[light-gray] (-2,-1) -- (-0.5, 0.5) --(1.5,-1.5) -- (1,-2) -- (0, -1) -- (-0.5,-1.5) -- (-1,-1) -- (-1.5,-1.5);
    \fill[light-gray] (4.5,-0.5) -- (6, 1) --(6.5,0.5) -- (5,-1);
    \fill[light-gray] (7,-1) -- (8, -2) --(7.5,-2.5) -- (6.5,-1.5);
    \node (0)  at (-1, 1.3) {$x$};
    \node (0)  at (-1, -0.7) {$a$};
    \node (0)  at (0.5, 1.8) {$y$};
    \node (0)  at (0, -0.7) {$b$};
    \node (0)  at (-0.03, 0.48) {$x\!\sqcap\! y$};
    \node (0)  at (-1, 1) {$\bullet$};
    \node (0)  at (-0.5, 0.5) {$\bullet$};
    \node (0)  at (-1, -1) {$\bullet$};
    \node (0)  at (0.5, 1.5) {$\bullet$};
    \node (0)  at (0, -1) {$\bullet$};

    \node (0)  at (6, 1.3) {$x$};
    \node (0)  at (6, -0.7) {$a$};
    \node (0)  at (7.5, 1.8) {$y$};
    \node (0)  at (7, -0.7) {$b$};
    \node (0)  at (6, 1) {$\bullet$};
    \node (0)  at (6, -1) {$\bullet$};
    \node (0)  at (7.5, 1.5) {$\bullet$};
    \node (0)  at (7, -1) {$\bullet$};
\path
	(-2.5,-0.5) edge (-1,1)
	(-1,1) edge (1.5,-1.5)
	(-1,-1) edge (-1.5,-1.5)
	(-1,-1) edge (0.5,-2.5)
	(0,-1) edge (-1,-2)
	(0,-1) edge (1,-2)
	(0.5,1.5) edge (2.5,-0.5)
	(0.5,1.5) edge (-2,-1)

	(4.5,-0.5) edge (6,1)
	(6,1) edge (8.5,-1.5)
	(6,-1) edge (5.5,-1.5)
	(6,-1) edge (7.5,-2.5)
	(7,-1) edge (6,-2)
	(7,-1) edge (8,-2)
	(7.5,1.5) edge (9.5,-0.5)
	(7.5,1.5) edge (5,-1)
	
	(-2.5,-0.5) edge[ultra thin,dashed] (0.5,2.5)
	(-2,-1) edge[ultra thin,dashed] (1,2)
	(-1.5,-1.5) edge[ultra thin,dashed] (1.5,1.5)
	(-1,-2) edge[ultra thin,dashed] (2,1)
	(-0.5,-2.5) edge[ultra thin,dashed] (2.5,0.5)
	(0,-3) edge[ultra thin,dashed] (3, 0)
	(0,-3) edge[ultra thin,dashed] (-3,0)
	(0.5,-2.5) edge[ultra thin,dashed] (-2.5,0.5)
	(1,-2) edge[ultra thin,dashed] (-2,1)
	(1.5,-1.5) edge[ultra thin,dashed] (-1.5,1.5)
	(2,-1) edge[ultra thin,dashed] (-1,2)
	(2.5,-0.5) edge[ultra thin,dashed] (-0.5,2.5)

	(4.5,-0.5) edge[ultra thin,dashed] (7.5,2.5)
	(5,-1) edge[ultra thin,dashed] (8,2)
	(5.5,-1.5) edge[ultra thin,dashed] (8.5,1.5)
	(6,-2) edge[ultra thin,dashed] (9,1)
	(6.5,-2.5) edge[ultra thin,dashed] (9.5,0.5)
	(7,-3) edge[ultra thin,dashed] (10,0)
	(7,-3) edge[ultra thin,dashed] (4,0)
	(7.5,-2.5) edge[ultra thin,dashed] (4.5,0.5)
	(8,-2) edge[ultra thin,dashed] (5,1)
	(8.5,-1.5) edge[ultra thin,dashed] (5.5,1.5)
	(9,-1) edge[ultra thin,dashed] (6,2)
	(9.5,-0.5) edge[ultra thin,dashed] (6.5,2.5);
\end{tikzpicture}
\vspace*{-0.4cm}
\caption{Intersection (left) and difference (right) of two pseudo-closures of pseudo-elements.}
\vspace*{-0.4cm}
\label{fig:op_pa}
\end{figure}
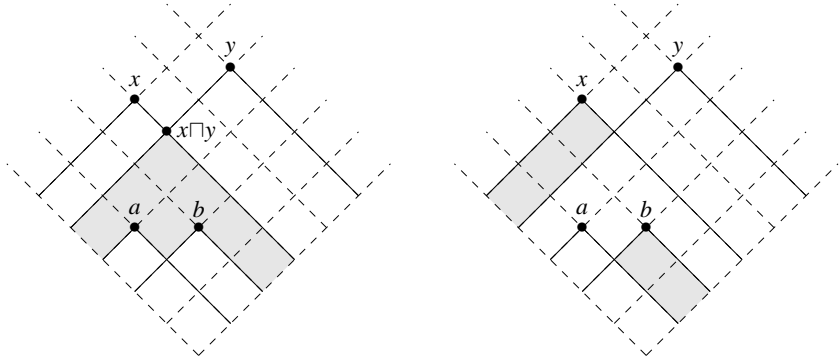

\medskip
\noindent The following example illustrates the second and third statements of the previous proposition.
\begin{example}
Let $(S, \preceq)$ be a lower semilattice and let $(x, \{a\})$ and $(y, \{b\})$, with $x, y, a, b \in S$, be two pseudo-elements as depicted in  Figure~\ref{fig:op_pa}. We have $\pseudclos(x, \{a\})$ $\cap$ $\pseudclos(y, \{b\}) =$ $\pseudclos(x\sqcap y, \{a,b\})$. We also have $\pseudclos(x, \{a\})$ $\setminus$ $\pseudclos(y, \{b\}) =$ $\pseudclos\{(x, \{y\}\antunion\{a\}), (x \sqcap b, \{a\})\} =$ $\pseudclos\{(x, \{y\}), (b, \{a\})\}$. Note that $(x, \{y\})$ and $(b, \{a\})$ are not in canonical form. The canonical representation of $\pseudclos(x, \{y\})$ (resp. $\pseudclos (b, \{a\})$) is given by $(x, \{x\sqcap y\})$ (resp. $(b, \{b\sqcap a\})$).
\end{example}

\section{Pseudo-antichain based algorithms}\label{sec:paalgo}
In this section, we propose a pseudo-antichain based version of the symblicit algorithm described in Section \ref{sec:si}  for solving the SSP and EMP problems for monotonic MDPs. 

\subsection{Operator $\Pre_{\actionO,\actionI}$}
We begin by presenting a new operator $\Pre_{\actionO,\actionI}$ that is useful for our algorithms.
Let $M_\preceq = (S, \ActionsO, \ActionsI, \edges, \distr)$ be a monotonic MDP. 
%
%
Given $L \subseteq S$, $\actionO \in \ActionsO$ and $\actionI \in \ActionsI$, we denote by $\Pre_{\actionO,\actionI}(L)$ the set of states that reach~$L$ by $\actionO,\actionI$ in $\monMDP$, that is 
$$\Pre_{\actionO,\actionI}(L) = \{s\in S \mid \edges(s,\actionO)(\actionI) \in L \}.$$ 
The elements of $\Pre_{\actionO,\actionI}(L)$ are called \textit{predecessors} of $L$ for $\actionO,\actionI$ in $\monMDP$.  This operator has the nice following properties: (1) if $L$ is closed, then $\Pre_{\actionO,\actionI}(L)$ is closed, and (2) for all sets $L_1$, $L_2$ and for all $\cdot~\in~\{\cup, \cap, \setminus\}$, $\Pre_{\actionO,\actionI}(L_1 \cdot L_2) = \Pre_{\actionO,\actionI}(L_1) \cdot \Pre_{\actionO,\actionI}(L_2)$. Moreover, we have:

%

\begin{proposition} \label{prop:anti}
\label{prop:Presigma}
Let $(x, \alpha)$ be a pseudo-element with $x \in S$ and $\alpha \subseteq S$. Let $A = \{(x_i, \alpha_i) \ | \ i \in I\}$ be a pseudo-antichain with $x_i \in S$ and $\alpha_i \subseteq S$ for all $i \in I$. Then, for all $\actionO \in \ActionsO$ and $\actionI \in \ActionsI$,
\begin{itemize}
	\item $\Pre_{\actionO,\actionI}(\pseudclos(x, \alpha)) = \bigcup_{x' \in \lceil \Pre_{\actionO,\actionI}(\antclos\ \{x\})\rceil}\pseudclos(x', \lceil\Pre_{\actionO,\actionI}(\antclos\alpha)\rceil)$
	\item $\Pre_{\actionO,\actionI}(\pseudclos A) = \bigcup_{i \in I}\Pre_{\actionO,\actionI}(\pseudclos (x_i, \alpha_i))$
\end{itemize}
\end{proposition}

%

From Proposition~\ref{prop:Presigma}, we can efficiently compute pseudo-antichains w.r.t. the $\Pre_{\actionO,\actionI}$ operator if we have an efficient algorithm to compute antichains w.r.t. $\Pre_{\actionO,\actionI}$ (see the first statement). We make the following assumption that we can compute the predecessors of a closed set by only considering the antichain of its maximal elements. Together with Proposition~\ref{prop:anti}, it implies that the computation of $\Pre_{\actionO,\actionI}(\pseudclos A)$, for all pseudo-antichains $A$, does not need to treat the whole pseudo-closure $\pseudclos A$.

\begin{assumption} \label{atwo}
There exists an algorithm taking any state $x \in S$ as input and returning $\lceil \Pre_{\actionO,\actionI}(\antclos \{x\})\rceil$ as output.
\end{assumption}



\begin{remark}\label{rem:ass1}
Assumption~\ref{atwo} is a realistic and natural assumption when considering partially ordered state spaces. For instance, it holds for the two classes of MDPs considered in Section~\ref{sec:experiments} for which the given algorithm is straightforward. Assumptions in the same flavor are made in~\cite{DBLP:journals/tcs/FinkelS01} (see Definition 3.2).
\end{remark}

\subsection{Symbolic representations}  \label{subsec:symbrep}

We detail in this section the symbolic representations based on pseudo-antichains that we are going to use in our algorithms. Recall from Section~\ref{sec:pa} that PA-representations are not unique. For efficiency reasons, it will be necessary to work with PA-representations that are as \textit{compact} as possible, as suggested in the sequel.

\paragraph{Representation of the stochastic models.}
Let $\monMDP = (S, \ActionsO, \ActionsI, \edges, \distr)$ be a monotonic MDP  and $\monMC = (S, \ActionsI, \edges_{\lambda}, \distr_{\lambda})$ be the MC induced by a strategy $\lambda$. 
For algorithmic purposes, in addition to Assumption~\ref{atwo}, we make the following assumption\footnote{Remark~\ref{rem:ass1} also holds for Assumption~\ref{aone}.}  on~$\monMDP$.

\begin{assumption} \label{aone} There exists an algorithm taking as input any state $s \in S$ and actions $\actionO \in \enabledactions, \actionI \in \ActionsI$, and returning as output $\edges(s, \actionO)(\actionI)$  and $\distr(s, \actionO)(\actionI)$.
\end{assumption}


By definition of $\monMDP$, $S$ is a closed set, and can thus be represented by the pseudo-antichain $\{(x,\emptyset) \mid x \in \lceil S \rceil\}$. By Assumption~\ref{aone}, we have a \PA-representation of $\monMDP$, in the sense that $S$ is \PA-represented and we can compute $\edges(s, \actionO)(\actionI)$  and $\distr(s, \actionO)(\actionI)$ on demand. 

Given $\lambda$,
we denote by $\equivstrat$ the equivalence relation on $S$ such that $s \equivstrat s'$ iff $\lambda(s) = \lambda(s')$. We denote by $\partstrat$ the induced partition of $S$. Given a block $B \in \partstrat$, we denote by $\lambda(B)$ the unique action $\lambda(s)$, for all $s \in B$. As any set can be represented by a pseudo-antichain, each block of $\partstrat$ is  \PA-represented. Therefore by Assumption~\ref{aone}, we have a \PA-representation of $\monMC$.

\paragraph{Representation of a subset of goal states.}
Recall that a subset $G \subseteq S$ of goal states is required for the SSP problem. Our algorithm will manipulate $G$ when computing the set of proper states. A natural assumption is to require that $G$ is \textit{closed} (like $S$), as it is the case for the two classes of monotonic MDPs studied in Section~\ref{sec:experiments}. Under this assumption, we have a compact representation of $G$ as the one proposed above for $S$.   
Otherwise, we take for $G$ any \PA-representation.

\paragraph{Representation for $\distr$ and $\reward$.}
For the needs of our algorithm, we introduce symbolic representations for $\distr_\lambda$ and $\reward_\lambda$.
Similarly to $\equivstrat$,  let $\equivdistr$ be the equivalence relation on $S$ such that $s \equivdistr s'$ iff $\distr_\lambda(s) = \distr_\lambda(s')$. We denote by $\partdistr$ the induced partition, and for each block $B \in \partdistr$, by $\distr_\lambda(B)$  the unique probability distribution $\distr_\lambda(s)$, with $s \in B$. We use similar notations for the relation $\equivreward$ on $S$ such that $s \equivreward s'$ iff $\reward_\lambda(s) = \reward_\lambda(s')$.
Each block of $\partdistr$ and $\partreward$ is \PA-represented.

For each $\actionO \in \ActionsO$, we also need the next equivalence relations $\sim_{\distr,\sigma}$ and $\equivsigmarew$ on $S$, such that $s \sim_{\distr,\sigma} s'$ iff $\distr(s,\sigma) = \distr(s',\sigma)$, and that $s \equivsigmarew s'$ iff $\reward(s, \actionO) = \reward(s', \actionO)$. 
Recall that $\distr$ and $\reward$ are partial functions, there may thus exist one block in their corresponding relation gathering all states $s$ such that $\actionO \not\in \ActionsO_s$. Each block of the induced partitions $S_{\sim_{\distr,\sigma}}$ and $S_{\sim_{\reward,\sigma}}$ is PA-represented.

For the two classes of MDPs studied in Section~\ref{sec:experiments}, both functions $\distr$ and $\reward$ are independent of $S$. It follows that the previously described equivalence relations have only one or two blocks, leading to compact symbolic representations of these relations. 

\medskip
Now that the operator $\Pre_{\actionO,\actionI}$ and the used symbolic representations have been introduced, we come back to the steps\footnote{Due to lack of space, some steps are not detailed. See~\cite{2014arXiv1402.1076B} for a complete description.} of the symblicit approach of Section~\ref{sec:si} (see Algorithm~\ref{algo:symblicit}) and show how to derive a pseudo-antichain based algorithm.

\subsection{Bisimulation lumping} 

\paragraph{Algorithm~\textsc{Lump}.} Let $\monMDP = (S, \ActionsO, \ActionsI, \edges, \distr)$ be a monotonic MDP and $\monMC = (S, \ActionsI, \edges_{\lambda}, \distr_{\lambda})$ be the MC induced by a strategy $\lambda$.\footnote{Equivalently, with the usual definition of MCs, $\monMC = (S, \probmat_\lambda)$ with $\probmat_\lambda$ derived from $\edges_{\lambda}$ and $\distr_{\lambda}$ (see Section~\ref{sec:prelim}).} In Algorithm~\ref{algo:symblicit}, Algorithm~\textsc{Lump} is called to compute the largest bisimulation $\equivlump$ of the MC $\monMC$ (line 4 with $\lambda = \lambda_n$). 
This algorithm (see~\cite{DBLP:journals/ipl/DerisaviHS03}) first computes the initial partition $P = \partreward$ such that two states of the MC are in the same block iff they have the same cost. It then repeatedly splits blocks $B$ of $P$ according to their probability of reaching a given block $C$, for all $C$. It stops as soon as for all $B, B' \in P$ and $s, s' \in B$, $\probmat_\lambda(s, B') = \probmat_\lambda(s', B')$. The operation of splitting blocks is performed with Algorithm~\textsc{Split} (see Algorithm~\ref{algo:split}). Before describing it, we need a new operator $\PreMC$. Given $L \subseteq S$ and $\actionI \in \ActionsI$, we define 
\vspace*{-0.15cm}
$$\PreMC(L, \actionI) = \{s \in S \mid \edges_{\lambda}(s)(\actionI) \in L\}$$ 
\vspace*{-0.2cm}
as the set of states from which $L$ is reached by $\actionI$ in $\monMDP$ under the selection made by $\lambda$. 

\paragraph{Algorithm~\textsc{Split}.} Given two blocks $B, C \subseteq S$, Algorithm~\textsc{Split} splits $B$ into a partition $P$ composed of sub-blocks $B_1, \ldots, B_k$ according to the probability of reaching $C$, i.e. for all $s,s' \in B$, $s,s' \in B_l$ for some~$l$ iff $\probmat_\lambda(s,C) = \probmat_\lambda(s',C)$.
%
%
Given $\ActionsI = \{\actionI_1, \ldots, \actionI_m\}$, it computes intermediate partitions $P$ of $B$ such that at step~$i$, $B$ is split according to the probability of reaching $C$ when $\ActionsI$ is restricted to $\{\actionI_1, \ldots, \actionI_i\}$.

Initially, $\ActionsI$ is restricted to $\emptyset$, and the partition $P$ is composed of one block $B$ (see line 1).
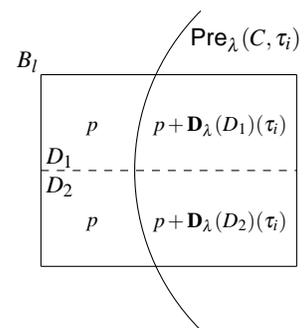
\begin{wrapfigure}{r}{50mm}
\vspace*{-0.3cm}
\centering
  \begin{tikzpicture}[-,>=stealth',auto,node distance=2.5cm,bend angle=45,scale=0.85,font=\footnotesize]
    \node (0)  at (0.8, 0.2) {$B_l$};
    \node (0)  at (1.3, -1.3) {$D_1$};
    \node (0)  at (1.3, -1.75) {$D_2$};
    \node (0)  at (4.2, 0.55) {$\PreMC(C,\actionI_i)$};
    \node (0)  at (1.8, -0.835) {{\scriptsize $p$}};
    \node (0)  at (1.8, -2.335) {{\scriptsize $p$}};
    \node (0)  at (3.8, -0.8) {{\scriptsize $p + \distr_\lambda(D_1)(\actionI_i)$}};
    \node (0)  at (3.8, -2.3) {{\scriptsize $p + \distr_\lambda(D_2)(\actionI_i)$}};
     \path
	(1,0) edge (5,0)
	(1,0) edge (1,-3)
	(1,-3) edge (5,-3)
	(5,-3) edge (5, 0)
	(1, -1.5) edge[dashed] (5, -1.5);
        \draw (3.5, 1) to[out=225,in=135] node {} (3.5,-4);
\end{tikzpicture}
\vspace*{-0.4cm}
\caption{Step $i$ of Algorithm~\ref{algo:split} on a block $B_l$.}
\label{fig:intuitionlump}
\vspace*{-0.3cm}
\end{wrapfigure}
At step $i$ with $i \geq 1$, each block $B_l$ of $P$ computed at step $i-1$ is split into several sub-blocks according to its intersection with $\PreMC(C,\actionI_i)$ and each $D \in \partdistr$. We take into account intersections with $D \in \partdistr$ in a way to know which stochastic function $\distr_\lambda(D)$ is associated with the states we are considering. Suppose that at step $i-1$ the probability for any state of block $B_l$ of reaching $C$ is $p$. Then at step $i$, it is equal to $p + \distr_\lambda(D)(\actionI_i)$ if this state belongs to $D \cap \PreMC(C,\actionI_i)$, with $D \in \partdistr$, and to $p$ if it does not belong to $\PreMC(C,\actionI_i)$ (lines 5-7). See Figure~\ref{fig:intuitionlump} for intuition. Notice that some newly created sub-blocks could have the same probability, they are therefore merged.

The intermediate partitions $P$ (or $P_\textnormal{new}$) are represented by hash tables for efficiency reasons:  each entry $(p, \val)$ is stored as $P[p] = \val$ such that $\val$ is the set of states that reach $C$ with probability $p$.
Algorithm~\textsc{InitTable} is called to initialize a new partition $P_\textnormal{new}$ from a previous partition $P$ and symbol $\actionI_i$: $P_\textnormal{new}[p] := \emptyset$ and $P_\textnormal{new}[p + \distr_\lambda(D)(\actionI_i)] := \emptyset$, for all $D \in \partdistr$ and all $(p, \val)$ in $P$. Algorithm \textsc{RemoveEmptyBlocks}$(P)$ removes from $P$ each pair $(p, block)$ with $block = \emptyset$.

\paragraph{Pseudo-antichain approach.} Notice that we have a pseudo-antichain version of Algorithm~\textsc{Lump} as soon as the given blocks $B$ and $C$ are PA-represented. Indeed, this algorithm uses boolean operations and the $\PreMC$ operator which can be computed as $\PreMC(C, \actionI)$ $=$ $\bigcup \big\{ \Pre_{\actionO,\actionI}(C)\ \cap B \mid \actionO \in \ActionsO, B \in \partstrat, \lambda(B) = \actionO \big\}$.
By Propositions~\ref{prop:opantichains}, \ref{prop:BooleanOp}-\ref{prop:Presigma} and Assumptions~\ref{atwo}-\ref{aone}, all these operations can be performed efficiently on pseudo-closures of pseudo-antichains, by limiting the computations to the related pseudo-antichains.

\begin{figure*}[t]
\vspace{-1cm}
 \begin{tabular}{lcr}
 \begin{minipage}[t]{.57\textwidth}
\begin{algorithm}[H]
\caption{{\footnotesize\textsc{Split}$(B, C, \lambda)$}}
{\scriptsize
\begin{algorithmic}[1] \label{algo:split}
\STATE $P[0] := B$
\FOR{$i$ in $[1, m]$}
	\STATE $P_\textnormal{new}  :=$ \textsc{InitTable}$(P,\actionI_i)$
	\FORALL{$(p, \val)$ in $P$}
		\STATE $P_\textnormal{new}[p] := P_\textnormal{new}[p] \cup (\val \diff \PreMC(C, \actionI_i))$
		\FORALL{$D \in \partdistr$}
			\STATE $P_\textnormal{new}[p + \distr_\lambda(D)(\actionI_i)] := P_\textnormal{new}[p + \distr_\lambda(D)(\actionI_i)]\cup (\val \cap D \cap \PreMC(C, \actionI_i))$
		\ENDFOR
	\ENDFOR
	\STATE $P:=$ \textsc{RemoveEmptyBlocks}$(P_\textnormal{new})$ 
	\ENDFOR
\STATE return $P$
\end{algorithmic}
}
\end{algorithm}
 \end{minipage}
 &&
 \begin{minipage}[t]{.34\textwidth}
 \begin{algorithm}[H]
\caption{{\footnotesize\textsc{ImproveStrat$({\cal L}, S_{\sim_{\lambda}})$}}}
{\scriptsize
\begin{algorithmic}[1] \label{algo:improvestrat}
\FOR{$C \in {\cal L}$}
	\STATE $S_{\equivstratp} := \emptyset$
	\FOR{$B \in S_{\equivstrat}$}
		\IF{$B \cap C \neq \emptyset$}
			\STATE $S_{\equivstratp} := S_{\equivstratp} \cup \{B\cap C, B\diff C\}$
		\ELSE
			\STATE $S_{\equivstratp} := S_{\equivstratp} \cup B$
		\ENDIF
	\ENDFOR
	\STATE $S_{\equivstrat} := S_{\equivstratp}$
\ENDFOR
\STATE return $S_{\equivstratp}$
\end{algorithmic}
}
 \end{algorithm}
 \end{minipage}
 \end{tabular}
\end{figure*}

%

\subsection{Improving strategies}

Given an MDP $\monMDP$ with cost function $\reward$ and the MC $\monMC$ induced by a strategy $\lambda$, we now present a pseudo-antichain based algorithm to improve strategy $\lambda$ for the SSP (see line 8 of Algorithm~\ref{algo:symblicit}). 
Recall that for all $s \in S$, we have to compute the set $\widehat{\enabledactions}$ of actions $\actionO \in \enabledactions$ that minimize the expression $\las = \reward(s, \actionO) + \sum_{s' \in S}\probmat(s, \actionO, s') \cdot v(s')$, and then we improve the strategy based on the computed $\widehat{\enabledactions}$ (see Algorithm~\ref{algo:SSPSI} with $v = v_n$ and $\lambda = \lambda_n$). We proceed in two steps: (1) we compute for all $\actionO \in \ActionsO$, an equivalence relation $\equiva$ such that the value $\las$ is constant on each block of the relation, (2) we use the relations $\equiva$, with $\actionO \in \ActionsO$, to improve the strategy.

\paragraph{Computing value $l_{\actionO}$.}
Let $\actionO \in \ActionsO$ be a fixed action. 
We are looking for an equivalence relation $\equiva$ on the set $\enabledstates$ of states where action $\actionO$ is enabled, such that
\begin{equation*}
\forall s, s' \in \enabledstates : s \equiva s' \Rightarrow \las = \lasp.
\end{equation*}
Given $\equivlump$ the largest bisimulation for $\monMC$ and the induced partition $\partlump$, we have $\las = \reward(s, \actionO) + \sum_{C \in \partlump} \probmat(s, \actionO, C) \cdot v(C)$ for each $s \in \enabledstates$,
since the value $v$ is constant on each block $C$. Therefore to get relation $\equiva$, it is enough to have $s \equiva s' \Rightarrow \reward(s, \actionO) = \reward(s', \actionO)$ and $\probmat(s, \actionO, C) = \probmat(s', \actionO, C), \forall C \in \partlump.$

We proceed by defining the following equivalence relations on $\enabledstates$. For the cost part, we use relation $\equivsigmarew$ defined in Section~\ref{subsec:symbrep}. For the probabilities part, for each block $C$ of $\partlump$, we define relation $\equivc$ such that $s \equivc s'$ iff $\probmat(s, \actionO, C) = \probmat(s', \actionO, C)$. The required relation $\equiva$ on $\enabledstates$ is then defined as the relation
$$\equiva ~=\ \ \ \equivsigmarew \cap \bigcap_{C \in \partlump}\!\!\equivc~=\ \ \ \equivsigmarew \cap \equivsigmaprob$$ 
Relation $\equiva$ induces a partition of $\enabledstates$ that we denote $\parta$. For each block $D \in\ \parta$, we denote by $\lad$ the unique value $\las$, for $s \in D$. 

Let us now explain how to compute $\equiva$ with a pseudo-antichain based approach. (1) The set $\enabledstates$ is obtained as $\enabledstates = \Pre_{\actionO,\actionI}(S)$ with $\actionI$ an arbitrary action of $\ActionsI$.
(2) Each relation $\equivc$ is the output returned by \textsc{Split}$(\enabledstates, C, \lambda)$ where $\lambda$ is defined on $\enabledstates$ by $\lambda(s) = \actionO$ for all $s \in \enabledstates$\footnote{As Algorithm~\textsc{Split} only works on $\enabledstates$, it is not a problem if $\lambda$ is not defined on $S \diff \enabledstates$.} (see Algorithm~\ref{algo:split}). (3) Let us detail a way to compute $\equivsigmaprob$ from $\equivc$, for all $C \in \partlump$. Let $\partc = \{B_{C,1}, B_{C,2}, \dots, B_{C,k_C}\}$ be the partition of $\enabledstates$ induced by $\equivc$. For each $B_{C,i} \in \partc$, we denote by $\probmat(B_{C,i}, \actionO, C)$ the unique value $\probmat(s, \actionO, C)$, for all $s \in B_{C,i}$. Then, computing a block $D$ of $\equivsigmaprob$ consists in picking, for all $C \in \partlump$, one block $D_C$ among $B_{C,1}, B_{C,2}, \dots, B_{C,k_C}$, such that the intersection $D = \bigcap_{C \in \partlump} D_C$ is non empty. As $\sum_{s' \in S}\probmat(s, \actionO, s') = 1$, if $D \neq \emptyset$, then $\sum_{C \in \partlump}\probmat(D_C, \actionO, C) = 1$. (4) Finally $\equiva$ is the intersection of $\equivsigmarew$ and~$\equivsigmaprob$.

\paragraph{Improving the strategy.} 

We now propose a pseudo-antichain based algorithm for improving strategy $\lambda$ by using relations $\equivlump$, $\equivstrat$, and $\equiva$, $\forall \actionO \in \ActionsO$ (see Algorithm~\ref{algo:improvestrat}).
We first compute for all $\actionO \in \ActionsO$, the equivalence relation $\equivalump\ =\ \equiva \cap \equivlump$ on $\enabledstates$.
Given $B \in \partalump$, we denote by $\laB$ the unique value $\las$ and by $v(B)$ the unique value $v(s)$, for all $s \in B$.  Let $\actionO \in \ActionsO$, we denote by $\partprom \subseteq \partalump$ the set of blocks $C$ for which the value $v(C)$ is improved by setting $\lambda(C) = \actionO$, that is 
$$\partprom = \{C \in \partalump \mid \lac < v(C)\}.$$ 
We then compute an ordered global list $\cal L$ made of the blocks of all sets $\partprom$, for all $\actionO \in \ActionsO$. It is ordered according to the decreasing value $\lac$. In this way, when traversing $\cal L$, we have more and more promising blocks to decrease $v$. 

From input $\cal L$ and $\equivstrat$, Algorithm~\ref{algo:improvestrat} outputs an equivalence relation $\equivstratp$ for a new strategy $\lambda'$ improving $\lambda$.
Given $C \in \cal L$, suppose that $C$ comes from the relation $\equivalump$ ($\actionO$ is considered). For each $B \in S_{\equivstrat}$ with $B \cap C  \neq \emptyset$ (line 4),  we improve the strategy by setting $\lambda'(B\cap C) = \actionO$, while $\lambda'$ is kept unchanged for $B\diff C$.
Algorithm~\ref{algo:improvestrat} outputs a partition $S_{\equivstratp}$ such that $s \equivstratp s' \Rightarrow \lambda'(s) = \lambda'(s')$ for the improved strategy $\lambda'$. If necessary, for efficiency reasons, we can compute a coarser relation for the new strategy $\lambda'$ by gathering blocks $B_1, B_2$ of $S_{\equivstratp}$, for all $B_1, B_2$ such that $\lambda'(B_1) = \lambda'(B_2)$. 


\section{Experiments} \label{sec:experiments}
We present two applications of the symblicit algorithm of Section~\ref{sec:paalgo}, one for the SSP problem in the context of automated planning, and the other for the EMP problem in the context of \LTL synthesis. In both cases, we have a reduction, described in~\cite{2014arXiv1402.1076B}, to monotonic MDPs equipped with a natural partial order and that satisfy Assumptions~\ref{atwo} and~\ref{aone}.
Our experiments have been done on a Linux platform with a $3.2$GHz CPU (Intel Core i7), with a timeout of $10$ hours and a memory usage limited to $4$GB.

\subsection{Stochastic shortest path on STRIPSs}
\paragraph{Monotonic STRIPS.} A \textit{monotonic STRIPS (MS)} is a tuple $(P, I, M, O)$ where $P$ is a finite set of \textit{conditions} (i.e. propositional variables), $I \subseteq P$ is a subset of conditions that are initially true (all others are assumed to be false), $M \subseteq P$ specifies which conditions must be true in a goal state, and $O$ is a finite set of \textit{operators}. An operator $o \in O$ is a pair $(\gamma, (\alpha, \delta))$ where $\gamma \subseteq P$ is the \textit{guard} of $o$, that is, the set of conditions that must be true for $o$ to be executable, and $(\alpha, \delta)$ is the \textit{effect} of $o$, that is, $\alpha \subseteq P$ (resp. $\delta \subseteq P$) is the set of conditions that are made true (resp. false) by the execution of $o$.
\textit{Monotonic stochastic STRIPS (MSS)} are MSs extended with stochastic aspects as follows~\cite{blum2000probabilistic}. Each operator $o = (\gamma, \pi) \in O$ consists of a guard $\gamma$ as before, and an effect given as a probability distribution $\pi : 2^P\times 2^P \rightarrow [0,1]$ on the set of pairs $(\alpha, \delta)$.  If additionally, we have a \textit{cost function} $C : O \rightarrow \R_{> 0}$, then the problem of planning from MSSs is to minimize the expected truncated sum up to the set of goal states from the initial state.
One can check that MSSs naturally define monotonic MDPs on which our algorithm of Section~\ref{sec:paalgo} can be applied for solving the mentioned SSP problem.

\paragraph{Experiments.} 
Our implementation is available at \url{http://lit2.ulb.ac.be/STRIPSSolver/} together with the two benchmarks presented in this section. It is compared with the purely explicit\footnote{To the best of our knowledge, there is no tool implementing an MTBDD based symblicit algorithm for the SSP problem. However, a comparison with an MTBDD based symblicit algorithm is done in the second application for the EMP problem.} strategy iteration algorithm implemented in the development release 4.1.dev.r7712 of the tool \prism~\cite{KNP11}. This explicit implementation exists primarily to prototype new techniques and is thus not fully optimized~\cite{Parker}. While value iteration algorithms are usually efficient, they only compute approximations. As a consequence, for the sake of a fair comparison, we consider here only the performances of strategy iteration algorithms. 

In the benchmark $\sf{Monkey}$ inspired from~\cite{russell1995artificial}, a monkey has several items (boxes, stones, pieces of sticks$\ldots$) at its disposal to reach a bunch of bananas, with the condition that it has to assemble some pieces to get a stick. 
The monkey has multiple ways to build the stick with varying building times.
The operators of getting some items and the bananas are stochastic. For instance, the probability of getting the bananas varies according to the owned items. The benchmark is parameterized in the pair $(p,s)$ where $p$ is the number of pieces to build a stick, and $p \cdot s$ is the total number of pieces. 
%
\begin{table} [t]
	\caption{{\small Stochastic shortest path on two benchmarks of MSSs. The column \textit{example} gives the parameters $(s, p)$ (resp. $(c, d)$) for the $\sf{Monkey}$ (resp. $\sf{Moats~and~castles}$) benchmark. The column $\ETP_\lambda$ gives the expected truncated sum of the computed strategy, and $|M_{\cal S}|$ the number of states of the MDP.
For the pseudo-antichain based implementation ($\sf{PA}$), $\#$it is the number of iterations of the strategy iteration algorithm, $|\partlump|$ the maximum size of computed bisimulation quotients, and \textit{lump}, \textit{syst} and \textit{impr} the total times (in seconds) spent respectively for lumping, solving the linear systems and improving the strategies. For both implementations, \textit{total} is the total execution time (in seconds) and \textit{mem} the total memory consumption (in megabytes).}}
	\label{table:STRIPS}
	\centering
		\scriptsize
 		\begin{tabular}{|rr|r|r||r|r|r|r|r|r|r||r|r|r|r|r|r|r|r|r|r|r|r|r|}
		\hline
		 & & & & \multicolumn{7}{|c||}{$\sf{PA}$} & \multicolumn{2}{|c|}{$\sf{Explicit}$}\rule[1pt]{0pt}{6pt}\\
		\multicolumn{2}{|r|}{example } & $\ETP_\lambda$  & $|M_{\cal S}|$  & $\#$it  & $|\partlump|$ & lump  &  syst  &  impr  &  total  & mem & total & mem\rule[-3pt]{0pt}{6pt}\\
\hline\hline
\multirow{9}{7pt}{\begin{sideways}$\sf{Monkey}$ \end{sideways}} 
%
%
%
& $(3,2)$ & $35.75$ & $4096$ & $4$ & $23$ & $0.09$ & $0.00$ & $0.07$  & $0.16$ & $16.0$ & $60.59$ & $1626$\rule[0pt]{0pt}{8pt}\\
& $(3,3)$ & $35.75$ & $65536$ & $5$ & $43$ & $1.14$ & $0.00$ & $0.43$  & $1.57$ & $17.3$ & $$ & $\memout$\rule[0pt]{0pt}{8pt}\\
& $(3,4)$ & $35.75$ & $1048576$ & $6$ & $57$ & $12.89$ & $0.01$ & $4.92$  & $17.83$ & $21.7$ & $$ & $\memout$\rule[0pt]{0pt}{8pt}\\
& $(3,5)$ & $36.00$ & $16777216$ & $7$ & $88$ & $208.33$ & $0.05$ & $63.73$  & $272.13$ & $37.5$ & $$ & $\memout$\rule[0pt]{0pt}{8pt}\\
\cline{2-13}
%
%
%
& $(5,2)$ & $35.75$ & $65536$ & $4$ & $31$ & $0.36$ & $0.00$ & $0.18$  & $0.54$ & $16.6$ & $20316.17$ & $2343$\rule[0pt]{0pt}{8pt}\\
& $(5,3)$ & $35.75$ & $4194304$ & $5$ & $56$ & $5.71$ & $0.02$ & $2.47$  & $8.20$ & $19.5$ & $$ & $\memout$\rule[0pt]{0pt}{8pt}\\
& $(5,4)$ & $35.75$ & $268435456$ & $6$ & $97$ & $95.49$ & $0.04$ & $101.27$  & $196.83$ & $31.3$ & $$ & $\memout$\rule[0pt]{0pt}{8pt}\\
& $(5,5)$ & $36.00$ & $17179869184$ & $7$ & $152$ & $1813.78$ & $0.08$ & $5284.31$  & $7098.40$ & $81.3$ & $$ & $\memout$\rule[0pt]{0pt}{8pt}\\
\hline\hline
\multirow{9}{7pt}{\begin{sideways}$\sf{Moats~and~castles}$\end{sideways}} 
& $(2,5)$ & $32.2222$ & $4096$ & $3$ & $49$ & $1.36$ & $0.00$ & $0.45$  & $1.82$ & $17.3$ & $133.66$ & $1202$\rule[0pt]{0pt}{8pt}\\
& $(2,6)$ & $32.2222$ & $16384$ & $3$ & $66$ & $9.71$ & $0.01$ & $1.95$  & $11.68$ & $19.3$ & $2966.80$ & $ 1706$\rule[0pt]{0pt}{8pt}\\
\cline{2-13}
& $(3,3)$ & $59.0000$ & $4096$ & $3$ & $84$ & $12.58$ & $0.03$ & $2.73$  & $15.35$ & $20.2$ & $149.64$ & $1205$\rule[0pt]{0pt}{8pt}\\
& $(3,4)$ & $52.0000$ & $32768$ & $3$ & $219$ & $129.17$ & $0.05$ & $21.56$  & $150.83$ & $30.7$ & $14660.69$ & $1611$\rule[0pt]{0pt}{8pt}\\
& $(3,5)$ & $48.3333$ & $262144$ & $3$ & $357$ & $658.86$ & $0.13$ & $81.08$  & $740.17$ & $49.1$ & $$ & $\memout$\rule[0pt]{0pt}{8pt}\\
& $(3,6)$ & $48.3333$ & $2097152$ & $3$ & $595$ & $10730.09$ & $0.42$ & $865.48$  & $11597.71$ & $145.8$ & $$ & $\memout$\rule[0pt]{0pt}{8pt}\\
\cline{2-13}
& $(4,2)$ & $96.8889$ & $4096$ & $3$ & $132$ & $31.61$ & $0.03$ & $12.06$  & $43.72$ & $26.5$ & $173.62$ & $1211$\rule[0pt]{0pt}{8pt}\\
& $(4,3)$ & $78.6667$ & $65536$ & $3$ & $464$ & $1376.94$ & $0.21$ & $217.06$  & $1594.48$ & $82.2$ & $$ & $\memout$\rule[0pt]{0pt}{8pt}\\
\hline
		\end{tabular}
		\normalsize
\end{table}

The benchmark ${\sf Moats~and~castles}$ is an adaptation of a benchmark of~\cite{DBLP:conf/aips/MajercikL98} as proposed in~\cite{blum2000probabilistic}\footnote{In~\cite{blum2000probabilistic}, the authors study the problem of maximizing the probability of reaching the goal within a given number of steps.}. The goal is to build a sand castle on the beach; a moat can be dug in a way to protect it. We consider up to $7$ discrete depths of moat. The operator of building the castle is stochastic: there is a strictly positive probability for the castle to be demolished by the waves. However, the deeper the moat is, the higher the probability of success is. 
To increase the difficulty of the problem, we consider building several castles, each one having its own moat. The benchmark is parameterized in the pair $(d,c)$ where $d$ is the number of depths of moat that can be dug, and $c$ is the numbers of castles to be built. 

Results are given in Table~\ref{table:STRIPS}. On those two benchmarks, the explicit implementation quickly runs out of memory when the state space of the MDP grows. Indeed, with this method, we were not able to solve MDPs with more than $65536$ states. On the other hand, the symblicit algorithm behaves well on large models: the memory consumption never exceeds $150$Mo and this even for MDPs with hundreds of millions of states\footnote{On our benchmarks, the value iteration algorithm of \prism~performs better than the strategy iteration one w.r.t. the run time and memory consumption. But it still consumes more memory than the pseudo-antichain based algorithm, and runs out of memory on several examples.}.


\subsection{Expected mean-payoff with \LTLMP synthesis}

\paragraph{\LTLMP synthesis problem.}
Let $\phi$ be an \LTL formula over a set $P = I\uplus O$. Let $\Sigma_P = 2^P$, $\Sigma_O = 2^O$ and $\Sigma_I = 2^I$, and $w : \Sigma_O \mapsto \Z$ be a weight function over $\Sigma_O$.
%
Consider the next infinite game between Player $O$ and Player $I$. At each turn $k$, Player $O$ gives $o_k \in \Sigma_O$ and Player $I$ responds by giving $i_k \in \Sigma_I$. The outcome of the game is the word $u = (o_0 \cup i_0)(o_1 \cup i_1)\dots \in \Sigma_P^\omega$. A \textit{value} $\mpval(u)$ is associated with $u$ such that
\begin{center}
$\mpval(u) =$
$\begin{cases} \liminf_{n \rightarrow \infty} \frac{1}{n} \sum_{k=0}^{n-1} w(o_k) \text{ if } u \models \phi\\  - \infty \text{ otherwise} \end{cases}$
\end{center}
i.e. $\mpval(u)$ is the mean-payoff value of $u$ if $u$ satisfies $\phi$, otherwise, it is $-\infty$.
%
%
Given a threshold value $\nu \in \Z$, the $\LTLMP$ \textit{realizability problem} asks whether Player~$O$ has a strategy against any strategy of Player $I$ such that $\mpval(u) \geq \nu$ for the produced outcome~$u$. 
The $\LTLMP$ \textit{synthesis problem} is to produce such a strategy for Player~$O$.

In~\cite{DBLP:journals/corr/abs-1210-3539,DBLP:conf/tacas/BohyBFR13}, we propose an antichain based algorithm for solving the $\LTLMP$ realizability and synthesis problems, that is incremental in some parameters $(K, C)$, and uses a reduction to a two-player turn-based safety game $G$ (see~\cite{DBLP:journals/corr/abs-1210-3539} for details). This game restricted to the winning positions of Player $O$ is a representation of a subset $W$ (depending on $(K, C)$) of all winning strategies for Player $O$ for $\LTLMP$ realizability problem.
%
%
From this set $W$, we want to compute a strategy that behaves \textit{the best against a stochastic opponent}. Let $\probI : \Sigma_I \rightarrow \ ]0,1]$ be a probability distribution such that $\pi(i) > 0$ for all $i \in \Sigma_I$ (to make sense with the worst-case). By replacing Player $I$ by $\probI$ in $G$, we can derive a monotonic MDP $M_G$ equipped with a natural partial order, and computing the best strategy  among strategies in $W$ reduces to solving the EMP problem for the MDP $M_G$\footnote{More precisely, it reduces to the EMP problem where the objective is to maximize the expected mean-payoff (see footnote~\ref{fn:altobj}).}.

\paragraph{Experiments.} 
We have integrated the symblicit algorithm presented in Section~\ref{sec:paalgo} for the EMP problem into \acaciaplus\ ($\sf{v2.2}$)~\cite{DBLP:conf/cav/BohyBFJR12}, a tool for solving the $\LTLMP$ realizability and synthesis problems. The latest version of \acaciaplus~can be downloaded at \url{http://lit2.ulb.ac.be/acaciaplus/}, together with the examples considered in this section. 
We compared our implementation with an MTBDD based symblicit algorithm implemented in \prism~\cite{prismEMP} (in the sequel, our implementation is simply called \acaciaplus~whereas the other one is called \prism). 
%
In the used benchmark~\cite{DBLP:conf/tacas/BohyBFR13}, a server has to grant exclusive access to a resource to two clients.
We set a probability distribution such that requests of client~$1$ (probability $\frac{3}{5}$) are more likely to happen than requests of client~$2$ (probability $\frac{1}{5}$),
and the benchmark is parameterized in the threshold value $\nu$.

Results are given in Table~\ref{table:LTL}. Note that the number of states in the MDPs depends on the implementation. Indeed, for \prism, it is the number of reachable states of the MDP, denoted $|M_G^R|$, that is, the states that are actually taken into account by the algorithm, while for \acaciaplus, it is the total number of states since unlike \prism, our implementation does not prune unreachable states. For this application~ scenario, the ratio (number of reachable states)/(total number of states) is in general quite small\footnote{For all the MDPs considered in Table~\ref{table:STRIPS}, this ratio is $1$.}. On this benchmark, \prism~is faster that \acaciaplus~on large models, but \acaciaplus~is more efficient regarding the memory consumption and this in spite of considering the whole state space. 
Note that the surprisingly large amount of memory consumption of both implementations on small instances is due to Python libraries loaded in memory for \acaciaplus, and to the JVM and the CUDD package for \prism~\cite{DBLP:conf/hvc/JansenKOSZ07}.

\begin{table}[t]
	\caption{{\small Expected mean-payoff on the $\sf{Stochastic~shared~resource~arbiter}$ benchmark with $2$ clients. The column $\nu$ gives the threshold, $|M_G^R|$ the number of reachable states in the MDP, and all other columns have the same meaning as in Table~\ref{table:STRIPS}. The expected mean-payoff $\EMP_\lambda$ of the optimal strategy $\lambda$ for all the examples is $-0.130435$.}}	\label{table:LTL}
	\centering
	\scriptsize
 	\begin{tabular}{|r||r|r|r|r|r|r|r|r||r|r|r|r|r|r|r|r|r|r|r|r|r|r|}
		\hline
	  	& \multicolumn{8}{|c||}{\acaciaplus} & \multicolumn{3}{|c|}{\prism}\rule[0pt]{0pt}{7pt}\\
		$\nu~$ & $|M_G|$  & $\#$it  & $|\partlump|$ & lump  &  LS  &  impr  &  total  & mem & $|M_G^R| $ & total & mem\rule[0pt]{0pt}{6pt}\\
\hline
$-1.1$ & $5259$ & $2$ & $22$ & $0.12$ & $0.01$ & $0.02$  & $0.15$ & $17.4$ & $691$  & $0.50$ & $168.1$\rule[0pt]{0pt}{8pt}\\
$-1.04$ & $35750$ & $2$ & $52$ & $1.63$ & $0.02$ & $0.13$  & $1.79$ & $18.1$ & $3325$ & $2.06$ & $264.1$\rule[0pt]{0pt}{8pt}\\
$-1.02$ & $530299$ & $2$ & $102$ & $16.62$ & $0.11$ & $0.64$  & $17.39$ & $20.2$ & $11641$ & $7.33$ & $343.4$\rule[-0pt]{0pt}{8pt}\\
$-1.01$ & $4120599$ & $2$ & $202$ & $237.78$ & $0.50$ & $3.94$  & $242.30$ & $26.2$ & $43891$ & $31.52$ & $642.5$\rule[0pt]{0pt}{8pt}\\
$-1.004$ & $63251499$ & $2$ & $502$ & $7357.72$ & $5.68$ & $52.81$  & $7416.77$ & $60.5$ & $264391$ & $278.01$ & $2544.0$\rule[0pt]{0pt}{8pt}\\
$-1.003$ & $450012211$ & $2$ & $670$ & $23455.44$ & $12.72$ & $120.25$  & $23589.49$ & $93.6$ & $$ & $$ & $\memout$\rule[0pt]{0pt}{8pt}\\
\hline
	\end{tabular}
	\normalsize
\end{table}

\medskip
Finally, in the majority of experiments we performed for both the EMP and the SSP problems, we observe that most of the execution time of the pseudo-antichain based symblicit algorithms is spent for lumping. It is also the case for the MTBDD based symblicit algorithm~\cite{DBLP:conf/qest/WimmerBBHCHDT10}.

\section{Conclusion} \label{sec:conclusion}
We have presented the interesting class of monotonic MDPs, and the new data structure of pseudo-antichains. We have shown how monotonic MDPs can be exploited by symblicit algorithms using pseudo-antichains (instead of MTBDDs) for two quantitative settings: the expected mean-payoff and the stochastic shortest path. Those algorithms have been implemented, and we have reported promising experimental results for two applications coming from automated planning and \LTLMP synthesis. We are convinced that pseudo-antichains can be used in the design of efficient algorithms in other contexts like for instance model-checking or synthesis of non-stochastic models, as soon as a natural partial order can be exploited.

\medskip
\textbf{Acknowledgments.} We would like to thank Mickael Randour for his fruitful discussions, Marta Kwiatkowska, David Parker and Christian Von Essen for their help regarding the tool $\sf{PRISM}$, and Holger Hermanns and Ernst Moritz Hahn for sharing with us their prototypical implementation.

\bibliographystyle{eptcs}
\bibliography{biblio}

\end{document}